\documentclass[english,aps,prl,twocolumn, superscriptaddress,showpacs, amsmath]{revtex4-1}

\newcommand{\noise}{$\langle\delta R^2\rangle/ \langle R\rangle^2$ }

\newcommand{\nbse}{2H-NbSe$_2$ }

\usepackage{bm}
\usepackage[normalem]{ulem}
\usepackage[T1]{fontenc}
\usepackage[latin9]{inputenc}
\usepackage{color}
\usepackage{units}
\usepackage{amssymb}
\usepackage{graphicx}
\usepackage{esint}
\usepackage{bm}
\usepackage{natbib}
\usepackage{ulem}
\usepackage[colorlinks=true, citecolor=red]{hyperref}
\setcitestyle{journalcolor= blue}
\usepackage{babel}

\begin{document}

\title{Quantum phase transition in few-layer NbSe$_2$ probed through quantized conductance fluctuations}

\author{Hemanta Kumar Kundu}
\affiliation{Department of Physics, Indian Institute of Science, Bangalore 560012, India}
\author{Sujay Ray}
\affiliation{Department of Physics, Indian Institute of Science, Bangalore 560012, India}
\author{Kapildeb Dolui}
\affiliation{Department of Physics, Indian Institute of Science, Bangalore 560012, India}
\author{Vivas Bagwe}
\affiliation{Tata Institute of Fundamental Research, Mumbai 400005, India}
\author{Palash Roy Choudhury}
\affiliation{Mahindra Ecole Centrale, Hyderabad 500043, India}
\author{S. B. Krupanidhi}
\affiliation{Materials Research Centre, Indian Institute of Science, Bangalore 560012, India}
\author{Tanmoy Das}
\affiliation{Department of Physics, Indian Institute of Science, Bangalore 560012, India}
\author{Pratap Raychaudhuri}
\affiliation{Tata Institute of Fundamental Research, Mumbai 400005, India}
\author{Aveek Bid}
\email{aveek@iisc.ac.in}
\affiliation{Department of Physics, Indian Institute of Science, Bangalore 560012, India}

\begin{abstract}
We present the first observation of dynamically modulated quantum phase transition (QPT) between two distinct charge density wave (CDW) phases in 2-dimensional 2H-NbSe$_2$. There is recent spectroscopic evidence for the presence of these two quantum phases, but its evidence in bulk measurements remained elusive. We studied suspended, ultra-thin \nbse devices fabricated on piezoelectric substrates - with tunable flakes thickness, disorder level and strain. We find a surprising evolution of the conductance fluctuation spectra across the CDW temperature: the conductance fluctuates between two precise values, separated by a quantum of conductance. These quantized fluctuations disappear for disordered and on-substrate devices. With the help of mean-field calculations, these observations can be explained as to arise from dynamical phase transition between the two CDW states. To affirm this idea, we vary the lateral strain across the device via piezoelectric medium and map out the phase diagram near the quantum critical point (QCP).  The results resolve a long-standing mystery of the anomalously large spectroscopic gap in NbSe$_2$. 
\end{abstract}

\maketitle

Despite intensive research over several decades, charge density waves (CDW) continue to remain at the forefront of modern condensed matter physics \cite{RevModPhysGruner, PhysRevLett.108.036404,doi:10.1080/00018732.2012.719674}. CDW in quasi-one dimension is understood to arise from Peierls mechanism - an inherent instability of a coupled electron-phonon system which creates a gap in the  single-particle excitation spectrum leading to the emergence of a collective mode formed of electron-hole pairs \cite{peierls1991more}. 
\begin{figure*}[t]
	\begin{center}
		\includegraphics[width=\textwidth]{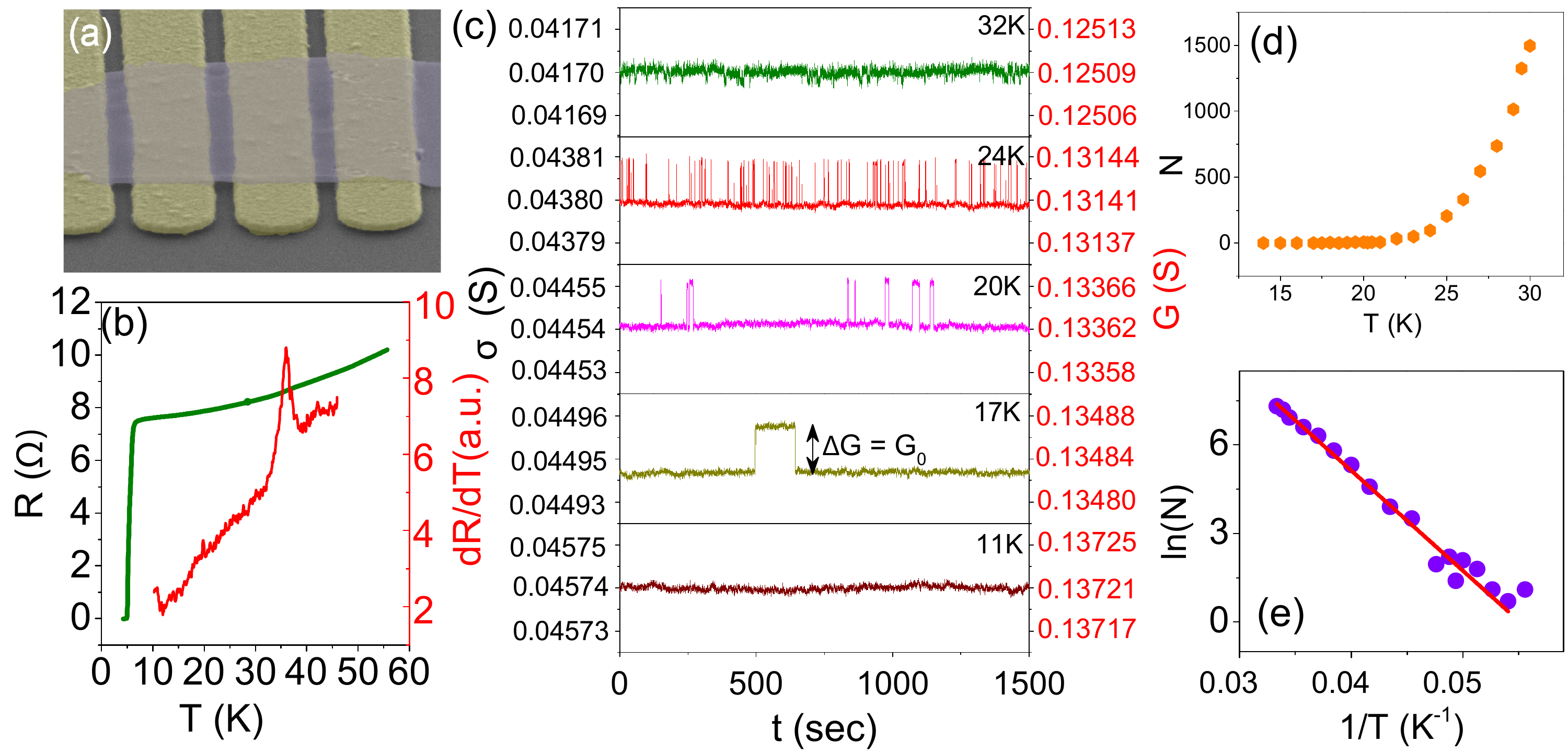}
		\small{\caption{ (a) False color SEM image of a typical suspended ultra-thin \nbse device. (b) Sample resistance $R$ (left axis) and its temperature derivative (right axis) as a function $T$ for device S1. (c) Plot of conductance per square, $\sigma$ (left-axis) and conductance (right-axis) measured at different $T$ for the same device. Note that with decreasing temperature, the frequency of the jumps reduce, although their amplitude remains unchanged. The length of the double headed arrow corresponds to $\Delta G= e^2/h$. 
				(d) Plot of the number of conductance switches, $N(T)$ over a thirty minute period versus temperature. (e) Fit of $N(T)$ to the Arrhenius equation. 				\label{fig:RT}}}
	\end{center}
\end{figure*}
In higher dimensions this electron-phonon interaction induced renormalization of the lattice wave vectors is often not enough to give rise to CDW \cite{PhysRevB.64.115116,PhysRevLett.118.106405,PhysRevB.77.165135, PhysRevLett.118.106405,PhysRevLett.114.037001, PhysRevLett.92.086401, PhysRevLett.95.117006, PhysRevLett.86.4382, PhysRevLett.99.146403, PhysRevLett.110.196404, PhysRevLett.105.176401}. One of the best known examples is 2H-NbSe$_2$, where the mechanism of CDW is still widely debated~\cite{ugeda2016characterization, PhysRevB.87.245111, borisenko2009two, johannes2006fermi, weber2011extended, flicker2015charge}. It has been suggested that the origin may lie in the strong momentum and orbital dependence of the electron-phonon coupling~\cite{johannes2006fermi, weber2011extended}. A natural consequence of this is the sensitivity of the CDW order to lattice perturbations. This has recently  been verified by Scanning Tunneling Microscopy (STM) measurements, which find the existence of 1Q striped quantum phase competing with the standard 3Q phase in locally strained regions~\cite{soumyanarayanan2013quantum}. The tri-directional 3Q phase respects the three-fold lattice symmetry and has a periodicity $Q\simeq 0.328 G_0$, where $G_0$ is the reciprocal lattice vector. The 1Q is a linear phase with a periodicity $Q\simeq (2/7)G_0$~ \cite{soumyanarayanan2013quantum}. Calculations indicate that, for  $T \ll T_{CDW}$, the system is very close to a quantum critical point  separating these two phases and any small perturbation, like local strain, can induce a quantum phase transition (QPT) between these two~\cite{ flicker2015charge,  flicker2015charge2, flicker2016charge3}. There are however, no direct experimental evidences of this QPT.


We probe for the possible existence of QPT in ultra-thin, suspended \nbse devices through time dependent conductance fluctuation spectroscopy~\cite{ghosh2004set}. We find that, for devices where the strain is dynamic, the electrical conductance fluctuates between two precise values separated by a quantum of conductance, with a well defined time scale. These fluctuations can be quenched either by damping out the strain fluctuations or by introducing lattice disorder into the system. We can control the transition between the two distinct quantum states by modulating the strain in devices fabricated on piezoelectric substrates. Through detailed calculations and analysis, we show that our observations are consistent with strain induced dynamic fluctuations between 3Q and 1Q  quantum phases in 2H-NbSe$_2$.  We also establish that the energy scale of $\sim$35~meV, often seen in spectroscopy studies in 2H-NbSe$_2$, is associated with the energy barrier separating the two CDW phases.

We study two classes of devices. The first class, which we call `on-substrate', is prepared on SiO$_2$/Si$^{++}$ substrates by mechanical exfoliation from bulk \nbse followed by standard electron beam lithography~\cite{ganguli2016disorder}. The second class of devices is suspended - few-layer \nbse flakes were mechanically exfoliated from bulk single crystals on silicone elastomer polydimethylsiloxane (PDMS) and transferred onto   Au electrodes pre-fabricated on either SiO$_2$/Si$^{++}$ or BaTiO$_3$/SrTiO$_3$ substrates. The aspect ratios (width/length ratio) of the samples were close to an integer, ranging from two to six. To study the effect of disorder, both these classes of devices are fabricated from multiple bulk \nbse crystals having a range of superconducting T$_c$ and residual resistivity ratios $\frac{R(300K)}{R(15K)}$)~\cite{ganguli2016disorder}.  Devices are also fabricated from bulk \nbse doped with cobalt to introduce disorder in a controlled manner. 
The devices range in thickness from bilayer to about 50 nm, as obtained both from optical contrast and AFM measurements [Supplementary Information]. A SEM image of a typical suspended device is shown in figure~\ref{fig:RT}(a).


Figure~\ref{fig:RT}(b) shows the evolution of the resistance, $R$ with temperature, $T$ of a suspended tri-layer device S1. The onset of CDW at $T_{CDW} \sim 35$~K is indicated by a peak in the dR/dT plot. The high value of the residual resistivity ratio, 8.5 and the relatively high superconducting $T_C$, 6~K indicate the defect free nature of the device. The time series of conductance fluctuations at different $T$ is plotted in figure~\ref{fig:RT}(c). For $T$ very close to $T_{CDW}$, the time series consists of random fluctuations about the average value, arising from the generic $1/f$ noise in the device. Below 30~K, we find the appearance of Random telegraphic noise (RTN) with the conductance fluctuating between two well defined levels separated by the quantum of conductance, $e^2/h$. 
The RTN persists right down to about 12~K below which superconducting fluctuations become dominant. The measurements are repeated on clean, suspended devices  of different flake thicknesses. It was seen that with increasing thickness, the magnitude of the conductance jumps increased, remaining in all cases close to an integer multiple of $e^2/h$ [Supplementary Materials Fig.~S5]. Figure~\ref{fig:RT}(d) shows a plot of the total number of switches over a period of 30~minutes at different temperatures. The switching statistics could  be well described by an Arrhenius function [Fig.~\ref{fig:RT}(e)].  The magnitude of the activation energy was found to be lie in the range $32 \pm 3$~meV in all such suspended, clean devices.

To probe in detail the statistics of the RTN, we performed low frequency resistance fluctuation spectroscopy at different temperature using a digital signal processing (DSP) based ac technique [Supplementary Information]. At each temperature the resistance fluctuations were recorded for 30 minutes. The resultant time-series of resistance fluctuations were digitally decimated and anti-aliased filtered. The power spectral density (PSD) of resistance fluctuations, $S_R(f)$ was calculated from this filtered time series using the method of Welch periodogram~\cite{ghosh2004set}. The $S_R(f)$ was subsequently integrated over the bandwidth of measurement to obtain the relative variance of resistance fluctuations: \noise=$\int{S_R(f)df}/\langle R\rangle^2$.

\begin{figure*}[t]
	\begin{center}
		\includegraphics[width=\textwidth]{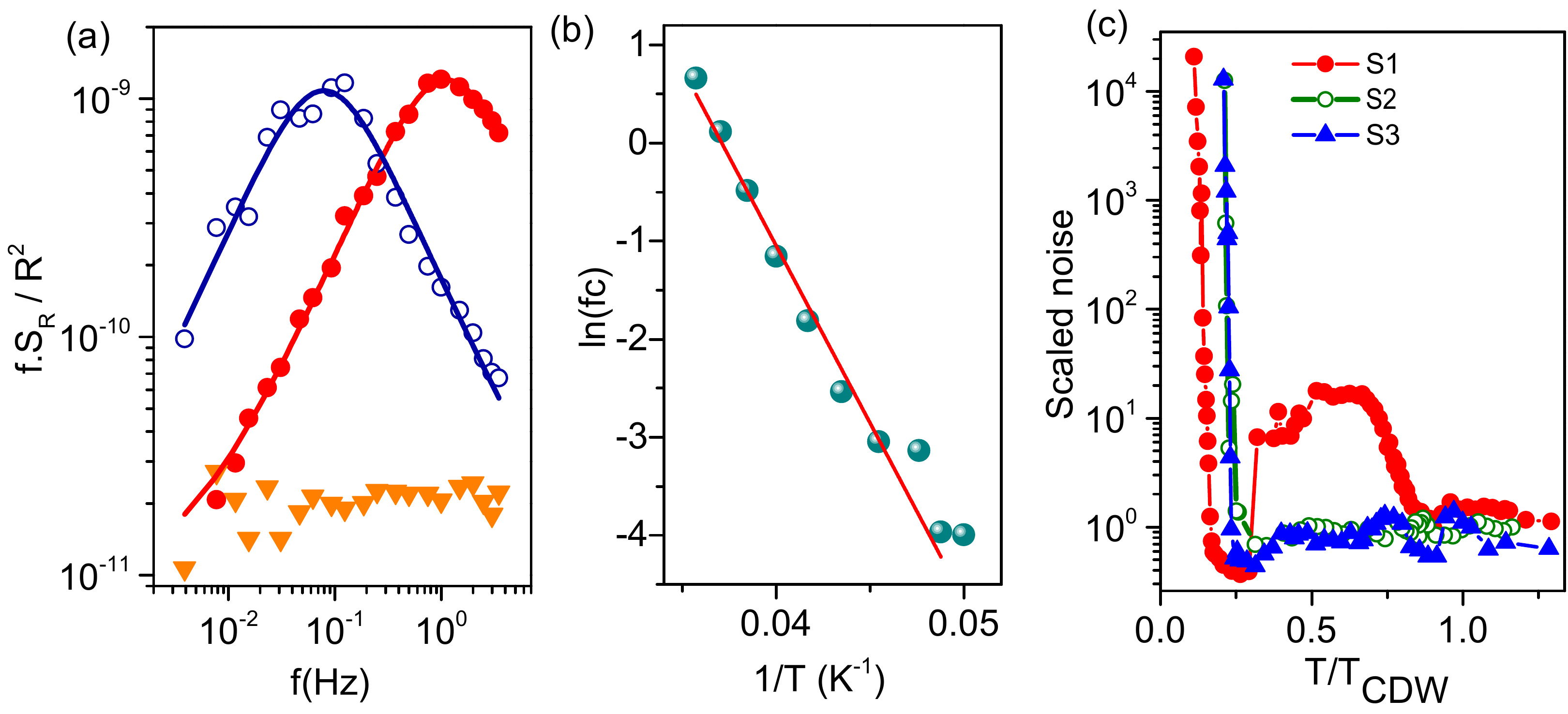}
		{\caption{(a) Scaled PSD of resistance fluctuations, $fS_R(f)/R^2$ versus $f$ at a few representative $T$ (open blue circle, filled red circle and inverted orange triangles correspond to data at 27~K, 23~K, 11~K respectively). The solid lines are fits to Eqn.~\ref{Eqn:2level}. (b) Plot of $f_C$ as a function of inverse temperature on a semi-log scale, the straight line is an Arrhenius fit to the data. (c) Plots of the relative variance of resistance fluctuations \noise (scaled by the value of \noise at $T=T_{CDW}$) versus $T/T_{CDW}$ for different classes of devices  - S1: clean tri-layer suspended device (red filled circle); S2: clean, approximately 25~nm thick substrated device (green open circle); and S3: Co-doped approximately 10 layer thick suspended device (blue triangle). 		
					\label{fig:fc}}}
	\end{center}
\end{figure*}

Figure~\ref{fig:fc}(a) shows the measured PSD at a few representative temperatures. We find that the PSD over the temperature window 12~K$<$T$<$30~K deviate significantly from $1/f$ nature, this $T$ range coinciding with that over which RTN was seen [Fig.~\ref{fig:RT}(c)]. The PSD of an RTN is a Lorentzian of corner frequency $f_C=1/\tau$ where $\tau$ is the time scale of the resistance switches between the two levels. Motivated by this, we analyzed the PSD data  using the relation:
\begin{equation}
\frac{S_R(f)}{R^2} = \frac{A}{f} + \frac{B.f_C}{f^2+f_C^2}
\label{Eqn:2level}
\end{equation}
The first term in Eqn.~\ref{Eqn:2level} represents the generic $1/f$ noise in the device, while the second term quantifies the contribution from a Lorentzian~\cite{fagerquist1989metastable}.  Constants A and B measure the relative strengths of the two terms and are derived from the fits to the experimental data [Fig.~\ref{fig:fc}(a)]. We find $f_C$ to be thermally activated, $f_C = f_0e^{-E_a/k_BT}$  [figure~\ref{fig:fc}(b)]. The value of the energy barrier is found to be $E_a = 35\pm$3~meV, which matches very well with that obtained from an analysis of the RTN jump statistics. 

The relative variance of resistance fluctuations $\langle\delta R^2\rangle/ \langle R\rangle^2$, normalized by its value at 60~K, is plotted in Fig. \ref{fig:fc}(c). The noise shows a broad peak over the temperature range  $0.3<T/T_{CDW}<0.9$ where RTN are present. We have verified that the additional contribution to the noise in this temperature range arises from the Lorentzian component in the PSD.  We find that over the temperature range where RTN are absent, the distribution of resistance fluctuations is Gaussian, as is expected for uncorrelated fluctuations. With decreasing $T$, \noise shoots up because of the onset of superconducting fluctuations. This has been seen before in many different superconducting systems and will not be discussed further  in this letter~\cite{koushik2013correlated,PhysRevB.94.085104,shi2016evidence}.

\begin{figure*}[!]
	\begin{center}
		\includegraphics[width=\textwidth]{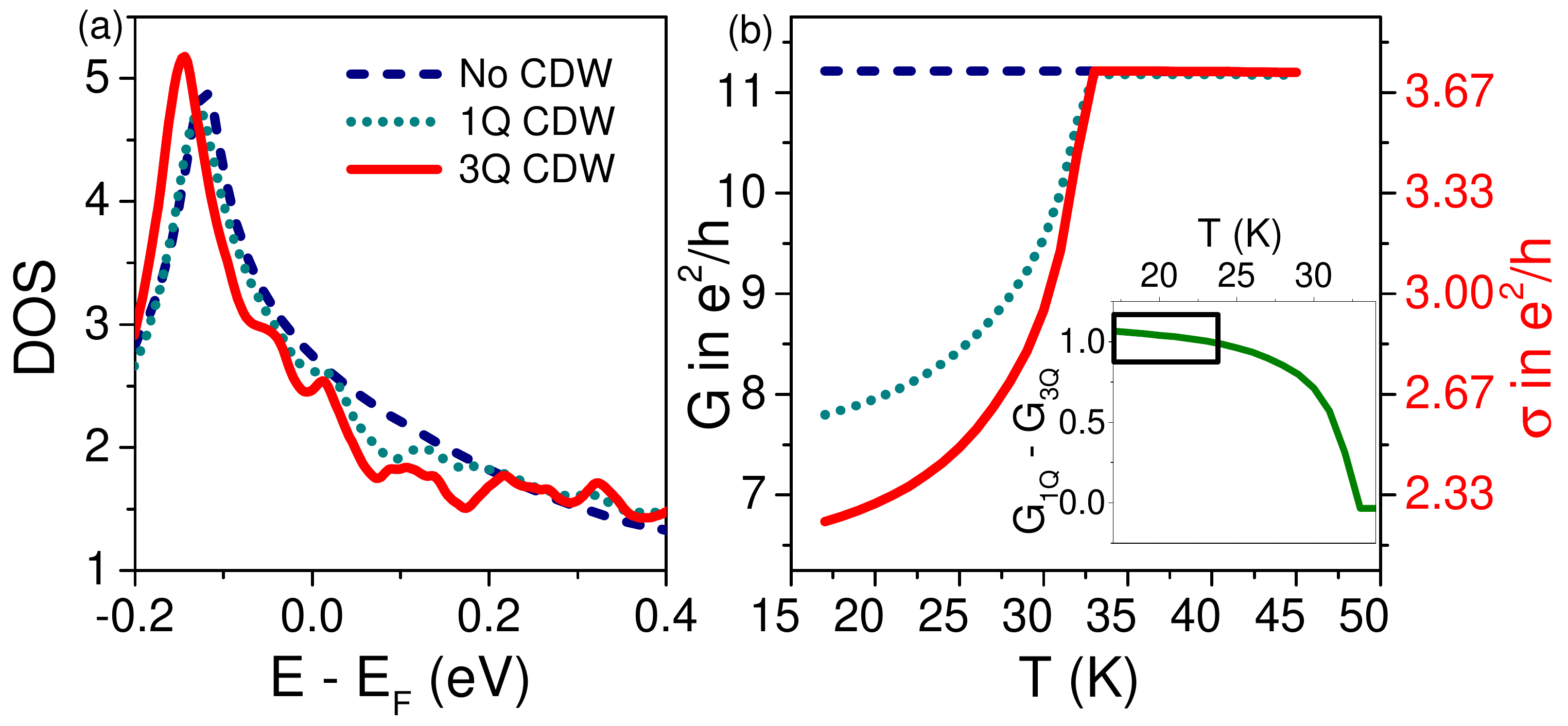}
				{\caption{ (a) Plot of the  calculated DOS in the two CDW phases. Also plotted for comparison is the DOS in the absence of CDW. (b) Temperature dependence of the computed conductance per square (right-axis) and conductance (left-axis) for the three cases. The inset shows the difference in conductance between the 3Q and the 1Q CDW phases over the temperature range where the RTN were observed. The box region is where the conductance fluctuations were experimentally found to be $e^2/h$.
				\label{fig:th}}}
	\end{center}
\end{figure*}

Turning now to the origin of these RTNs, we note that these can possibly arise, in CDW systems which have a single-particle energy gap at the Fermi level (e.g. NbSe$_3$ and TaS$_3$)~\cite{bloom1994discrete,marley1994temperature,bloom1994correlation}, due to the switching of the ground state of the system between pinned and sliding states. In some of these systems sharp noise peaks were observed even at values of electric fields lower than the threshold field for slippage of the CDW~\cite{pokrovskii1989solitary}.Our results differ from what was observed in these systems in two important aspects: (1) our measurements were performed under electric fields of magnitude few V/m which were at least two orders of magnitude smaller than the electric fields applied to observe RTN in these systems ~\cite{pokrovskii1989solitary, bloom1994discrete,marley1994temperature,bloom1994correlation}, and (2) unlike in NbSe$_3$ and TaS$_3$, the observed RTN in \nbse were independent of electric field~\cite{pokrovskii1989solitary}. However, unlike NbSe$_3$ and TaS$_3$, the CDW in \nbse does not slide. This is consistent with our observation that the RTN in \nbse were independent of electric field. This suggests that  RTN in \nbse  must have an origin distinct from those seen in gapped CDW systems like NbSe$_3$ and TaS$_3$.

There is a due concern that the observed RTN may arise due to the interplay of superconducting fluctuations above $T_c$ and CDW order. Measurements performed under perpendicular magnetic fields much higher than $H_{c2}$ of bulk \nbse do not have any effect on either the frequency or the amplitude of these two level fluctuations, ruling out this interpretation [Supplementary information]. We also considered the possibility that the RTN can arise due to the quantization of the  number of density waves along the perpendicular direction, as seen in some systems~\cite{kummamuru2008electrical,PhysRevLett.98.117206}. We ruled this out by noting that in \nbse the weak inter-layer van der waals interaction precludes the formation of any long range density waves perpendicular to the planes. This is supported by spectroscopic studies.

The most compelling explanation of the RTN we observe in \nbse is  phase fluctuations  between 1Q and 3Q phases. Earlier calculations~\cite{flicker2015charge2}, supported by the STM measurements~\cite{soumyanarayanan2013quantum}, demonstrated that the crossover between 3Q and a 1Q CDW phases at a given temperature can be induced by a strain as small as 0.1\%. Experiments show that suspended \nbse devices in contact with Au pads experience an average strain of about 0.1\% at low temperatures~\cite{PhysRevB.82.155432} which is sufficient to drive the system close to the boundary separating these two quantum phases~\cite{flicker2015charge2}. In such a suspended mesoscopic device, at a finite temperature, the strain dynamically fluctuates due to thermally enhanced mechanical vibrations. This fluctuating strain can lead to a dynamical phase transition from 3Q to 1Q and vice versa in \nbse at a fixed temperature. This would cause the conductance of the system to fluctuate between two well defined values if the conductivity of the two phases are different. We validate this conjecture through detailed Density-functional theory (DFT) based band structure calculations of the conductance in the two distinct quantum phases of 2H-NbSe$_2$.

We calculate the DC conductivity $\sigma$ in both the 3Q and 1Q CDW phases using a two-band model, relevant for this compound~\cite{PhysRevB.64.235119}. The non-interacting dispersions $\xi_{1k,2k}$ are directly deduced from the DFT calculations [Supplementary Information]. \footnote{\label{myfootnote} The wavevector of the CDW state is known to be ${\bf Q}_{\nu} \approx 1/3$ ${\bf G}^{\nu}_0$, where ${\bf G}^{\nu}_0$ are the three reciprocal lattice vectors, and $\nu=1, 2, 3$ in the 3Q phase. In the 1Q phase, only one of the ${\bf Q}_{\nu}$ values remain active along the CDW propagation direction (we take ${\bf Q}_1$)}. The CDW order parameters are introduced within the mean-field approximation:
\begin{eqnarray}
H &=& \sum_{i,{\bf k}}\Big[\xi_{i,{\bf k}} c^{\dag}_{i,{\bf k}}c_{i,{\bf k}} + \sum_{\nu}\big(\xi_{i,{\bf k}+{\bf Q}_{\nu}} c^{\dag}_{i,{\bf k}+{\bf Q}_{\nu}}c_{i,{\bf k}+{\bf Q}_{\nu}},\nonumber\\
&&\qquad\qquad\qquad\qquad +\Delta_{i,\nu} c^{\dag}_{i,{\bf k}}c_{j,{\bf k}+{\bf Q}_{\nu}}\big)  \Big]+ {\rm h.c.}.
\label{Ham}
\end{eqnarray}

\noindent Here the band index $i=1,2$, and the nesting index $\nu$ takes 3 values in the 3Q phase and 1 value in the 1Q phase. $c_{i,{\bf k}}$  is the annihilation operator for the electron in the $i^{\rm th}$-band at momentum ${\bf k}$. The mean-field CDW gap $\Delta_{i,\nu}$ is defined between the two bands. We obtain the quasiparticle energies $E_{i,{\bf k}}$ by exact diagonalization of the Hamiltonian in ~\eqref{Ham}, and there are four, and eight quasiparticle states in the 1Q and 3Q phases, respectively.  

\begin{figure*}[t!]
	\begin{center}
		\includegraphics[width=\textwidth]{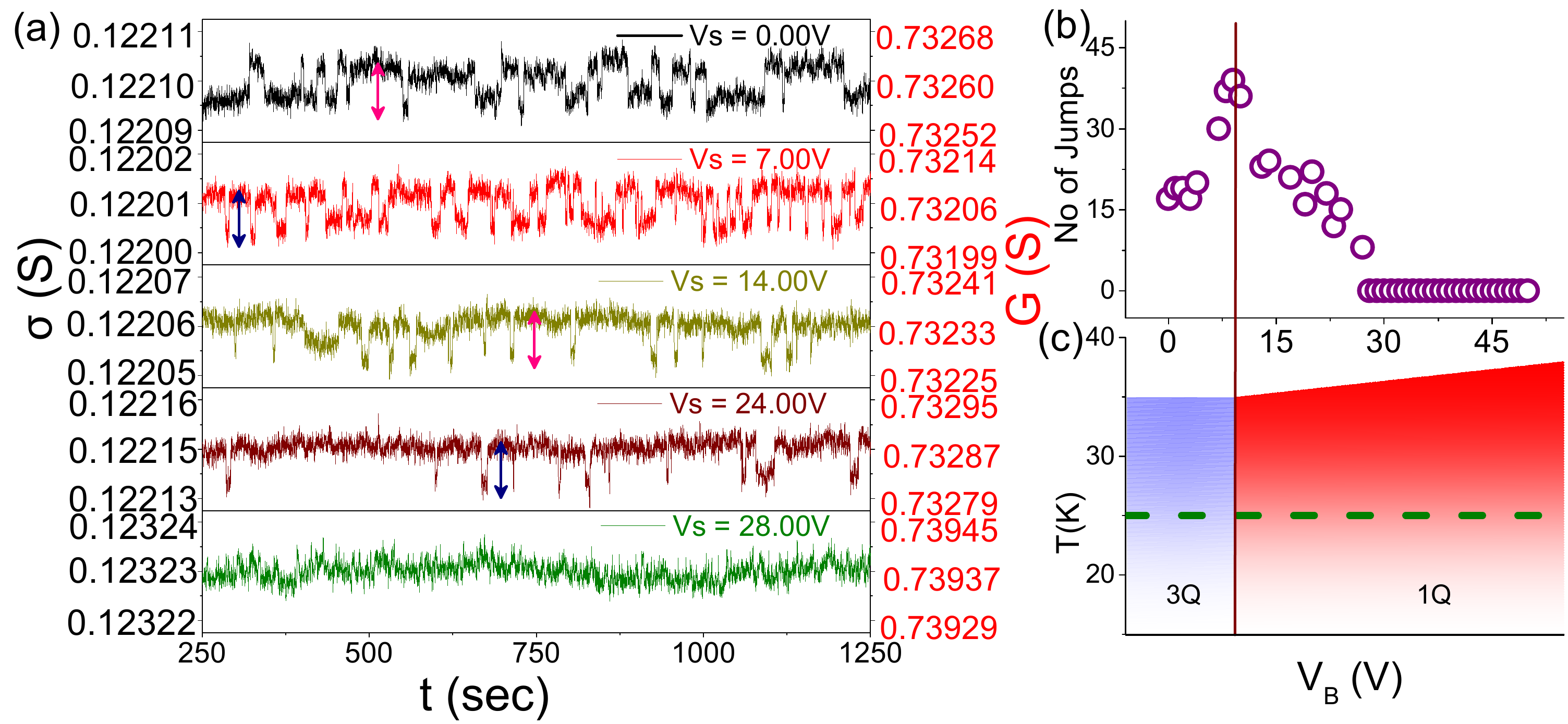}
		\small{\caption{(a) Plot of the time-series of conductance per square (left-axis) and conductance (right-axis) measured at T = 25~K for the suspended 10 layer thick \nbse device (S5) fabricated on BTO substrate. The numbers in the legend refer to the voltage $V_B$ across the substrate while the length of the double headed arrows correspond to $\Delta G= 2e^2/h$. (b) Dependence of the number of conductance jumps, measured over a period of thirty minutes on  the voltage $V_B$. 	  (c) 		Schematic phase diagram of the system in temperature-strain plane. The blue shaded region is in the 3Q phase while the pink shaded region is in the 1Q phase. The green dotted line represents the isotherm at 25~K along which the data presented in panels (a) and (b) were collected.	\label{fig:strain}}}
	\end{center}
\end{figure*} 

The conductivity of the two phases primarily depends on the CDW gap values $\Delta_{i,\nu}$, which are related to the CDW potential $V_{\nu}$ by $\Delta_{i,\nu}=V_{\nu}\sum_{i,{\bf k}}\frac{\Delta_{i,\nu}}{2E_{i,{\bf k}}}{\rm tanh}(\beta E_{i,{\bf k}}/2)$, with $\beta=1/k_BT$. The interaction $V_{\nu}$ arises from the electron-phonon coupling~\citep{flicker2015charge2} and is directly related to strain, and therefore it becomes directional dependent. A CDW phase arises along a direction $\nu$ when the corresponding strain induced potential exceeds the critical potential $V_{\nu}>V_c\sim2W$, where $W\sim 1.21 eV$ is the bandwidth. Since the present system reside in the vicinity of the critical point, $V_{\nu}\sim V_c$, and the phase diagram is very sensitive to strain. In the 3Q phase, all three $V_{\nu}>V_c$, while in 1Q phase, only $V_1>V_c$, and the rest are $<V_c$.  


In the mean-field state, we find a substantial suppression of the density of state (DOS) at $E_F$ in the 3Q phase, with a gap which is calculated to be $\Delta_0\sim$35~meV~[Fig.~\ref{fig:th}(a)]. However, in the 1Q phase, the spectral weight loss at $E_F$ is significantly less. These results are consistent with the STM data~\cite{soumyanarayanan2013quantum}. Therefore, we anticipate that the conductivity in the 3Q phase will be lower than that in the 1Q phase. 

We first calculate the conductivity $\sigma$ using the standard Kubo formula. We then  obtain the conductance G by normalizing the value of $\sigma$ with the dimensions of the present device ($G=\sigma \times$(width/length)). We assume  band independent gap values. For the ratio of $\Delta_{\rm 3Q}/\Delta_{\rm 1Q}=1.06$, we find that the difference in conductance between the two CDW phases, $\Delta G=G_{\rm 1Q}-G_{\rm 3Q}\sim e^2/h$, as seen experimentally. We also notice that over the temperature range $T=$17K-24K,  $\Delta G$ changes very little as the self-consistent gap remains essentially unchanged over this narrow temperature window [Fig.~\ref{fig:th}(b)]. This result is consistent with our experimental observations.  We do not have a microscopic understanding of why this quantity should be an integral multiple of $e^2/h$. This may require the inclusion of topological terms in the calculation which is beyond the scope of the present work.

If dynamical phase fluctuations between the two CDW phases is indeed responsible for the observed RTN, it should be possible to modulate the frequency of the conductance jumps by driving the system controllably between the two competing CDW phases. To test this hypothesis, suspended devices of few layer \nbse are fabricated on piezoelectric BaTiO$_3$/SrTiO$_3$ (BTO) substrates. In this device,  the strain across the device can be modulated by varying the voltage $V_{B}$ across the substrate.  Fig.~\ref{fig:strain}(a) shows the evolution of the conductance fluctuations with changing $V_B$ obtained for one such device at 25~K. At very low values of $V_B$ (strain), the frequency of the conductance jumps is low and the system is seen to spend statistically similar amounts of time in both the high  and low conductance states. With increasing $V_{B}$ (and consequently increasing strain across the device), the frequency of the conductance jumps initially increases and then decreases rapidly. However, the magnitude of the conductance jumps throughout this process remained quantized in units of $e^2/h$. Eventually, the conductance jumps  vanishes as the system stabilized in the higher conduction state [Fig.~\ref{fig:strain}(b)]. We note that in different sweep cycles in $V_B$ the RTN are not exactly reproducible. It is difficult at this stage to comment on whether this is due to inherent hysteresis in the piezoelectric response of BTO or if it indicates non-reversibility of the properties of NbSe$_2$.


These results can be understood as follows: with increasing strain \textit{via} $V_B$, the system approaches the phase boundary separating the 3Q and 1Q phases, leading to an increased probability of switching between the two states. Eventually, the system crosses the phase boundary and consequently, the switching frequency starts decreasing and finally vanishes as the system settles into the 1Q state. These measurements establish conclusively that, consistent with theoretical calculations, strain can drive the system to the higher conducting 1Q phase from lower conducting 3Q phase.  

As seen from STM measurements on substrated devices, local random strain due to lattice imperfections causes the system to spatially phase separate into an inhomogeneous mixture of 3Q and 1Q phases~\cite{soumyanarayanan2013quantum}. This local phase separation can not cause the measured conductance, which is a macroscopic global averaged property, to fluctuate dynamically between two well defined conductance levels  separated by the quantum of conductance.   To validate this conjecture, measurements were performed on \nbse devices of various thicknesses prepared on SiO$_2$/Si$^{++}$ substrates [Supplementary Information]. Although we observed clear CDW transition in this set of devices from resistivity measurements, no signature of RTN was seen in any of them.  The conductance fluctuations in these devices, at all temperatures $T>T_C$, consisted only of generic $1/f$ fluctuations arising from defect dynamics. The magnitude of noise \noise remained constant over the temperature range $T_{CDW}>T>T_C$ before showing the  sharp rise near superconducting transition [Fig. \ref{fig:fc}(c)]. 

To test the effect of disorder, measurements were performed on suspended devices exfoliated from  bulk \nbse crystals having low bulk T$_c$ and low residual resistivity ratio and from bulk \nbse crystals doped with 0.1\% Co. Atomic Force microscopy measurements showed that the rms surface roughness of the low T$_c$ flakes was about 3 times higher than that of the high T$_c$ flakes [Supplementary Information].  Although we observed dR/dT peak at ~35 K in these devices indicating the presence of CDW,  we did not observe RTN in any of them. The noise in these devices was similar to what was seen for substrated devices indicating the suppression of RTN due to disorder in the system~[Fig. \ref{fig:fc}(c)]. The absence of RTN in all the control experiments involving substrated, Co-doped and disordered suspended \nbse devices, as well as the insensitivity of the conductance fluctuations in clean suspended devices to high magnetic fields reinforces our interpretation of the origin of the observed RTN in clean suspended devices as lattice fluctuation mediated.

To conclude, in this letter we demonstrate controlled, strain induced phase transition between the 1Q and 3Q CDW phases  in suspended 2D 2H-NbSe$_2$. With this, we resolve a long standing question of finite temperature dynamic phase transition between two quantum phases of the CDW system. We show the energy scale of $\pm$ 35meV, seen repeatedly in spectroscopic measurements~\cite{borisenko2009two, soumyanarayanan2013quantum}, to be the barrier corresponding to 1Q-3Q phase transition. Our work establishes conductance fluctuation spectroscopy as a technique to probe phase co-existence and phase transitions in nanoscale systems and can thus be a step forward in the understanding of competing quantum phases in strongly correlated systems.

A.B. acknowledges financial support from Nanomission, DST, Govt. of India  project SR/NM/NS-35/2012; SERB, DST, Govt. of India  and Indo-French Centre for the Promotion of Advanced Recearch (CEFIPRA). H.K.K. thanks CSIR, MHRD, Govt. of India for financial support. TD acknowledgesthe financial support from the DST, India under the Start Up Research Grant (Young Scientist) [SERB No: YSS/2015/001286]. We acknowledge discussions with A.~V.~Mallik, B. F. Gao, D.~Nordlund, A.~N.~Pasupathy and Diptiman Sen.


\clearpage

\section*{Supplementary Information}
\renewcommand{\thesection}{S\arabic{section}}

\section{Effect of magnetic field on the observed RTN:} In many low-dimensional superconductors, superconducting fluctuations can persist at temperatures much higher than the mean field transition temperature $T_C$. In the case of our cleanest bulk samples, the measured $T_C$ was $\sim7.2~K$. Thus, there might be a concern that the two-level  conductance fluctuations observed by us in the clean suspended \nbse flakes at 15~K and above might have some contributions from superconducting phase/amplitude fluctuations. To rule out this possibility, we have studied the effect of magnetic field $B$ (with $B$ much larger than the critical field $B_C$) on the observed RTN. Fig.~\ref{fig:S1} shows plots of conductance fluctuation measured at 24~K at zero field and in the presence of an 8~T perpendicular field, the data are statistically identical showing that magnetic field had no discernible effect on the two-level conductance fluctuations.

\section{Resistance fluctuation spectroscopy:} To probe resistance fluctuation and its statistics, we used a standard 4-probe  digital signal processing (DSP) based ac noise measurement technique~\cite{ghosh2004set}. This technique allows us to simultaneously measure the background noise as well as sample noise. The device was biased by a constant ac current source, typical currents used during the measurement were 10 $\mu$A. A low-noise pre-amplifier (SR552) was used to couple the voltage across the device to dual channel digital lock-in-amplifier (LIA). The bias frequency of the LIA (228~Hz) was chosen to lie in the eye of the noise figure of the pre-amplifier to reduce the contribution of amplifier noise. The output of the LIA was digitized by a high speed 16 bit analog-to-digital conversion card and stored in computer. The complete data set for each run, containing $1.5 \times 10^{6}$ data points, was decimated and digitally filtered to eliminate the 50~Hz line frequency component. This filtered time series was then used to calculate the power spectral density (PSD) of voltage fluctuation S$_V$ over specified frequency window using the method of Welch Periodogram. The lower frequency limit $\sim 4$~mHz and the upper frequency limit $\sim4$~Hz were  limited by the ratio of the sample noise to background noise. The PSD of voltage fluctuation was converted to the PSD of resistance fluctuation $S_R(f)$ by the relation $S_R(f) = \frac{S_V (f)}{I^2}$ where $I$ is rms value of constant current used to bias the device. The measurement set up has been calibrated by thermal noise measurements on standard resistors to measure spectral density down to $S_V ~ 10^{-20} V^2 Hz^{-1}$. The measured thermal background noise on the \nbse devices were found to be bias independent and frequency independent; and the PSD matched the value of 4$k_BTR$, as expected from Johnson-Nyquist noise. The PSD of resistance fluctuation was subsequently integrated over the bandwidth of measurement to obtain the relative variance of resistance fluctuations: $$\frac{\langle\delta R^2\rangle}{\langle R^2\rangle} = \frac{1}{R^2}\int_{f_{min}}^{f_{max}} S_R(f)df $$

\section{Absence of RTN in on-substrate and disordered \nbse devices:}
To conclusively establish that the RTN seen by us in \nbse is present only in suspended clean devices, we fabricated on-substrate devices from the same high quality bulk \nbse from which the suspended devices showing RTN were exfoliated. In all our substrated device, we observed CDW transition with similar T$_{CDW}$ ~ 35 K but did not find signatures of RTN at any temperature [fig.~\ref{fig:Substrated_RT_TS}]. Similarly, suspended devices fabricated from bulk \nbse crystals having low superconducting $T_C$, despite undergoing a CDW transition at 35~K, did not show any RTN as shown in \ref{fig:Substrated_RT_TS}[d]. These control experiments confirm that the two-level conductance fluctuations seen by us are a property of clean suspended \nbse devices.
We have performed AFM measurements to map the topography of the flakes and their thicknesses. It was observed that flakes exfoliated from bulk crystal of lower T$_C$ had a much higher surface roughness ($\approx$ 3-4 times) than those from high quality bulk crystals - the AFM topography images are shown in fig.~\ref{fig:AFM}. The observation of RTN in all the different classes of samples measured is summarised in fig. \ref{fig:flowchart}. This flowchart shows that RTN is observed only in suspended devices made from disorder free flakes.

In Ref.~\cite{soumyanarayanan2013quantum} the two phases are found to coexist due to non-uniform local strains because of underlying defects -- these devices were all on substrate.  Since the thermodynamic phase of the system is well defined in the temperature-phase plane, for uniform strain the whole sample will undergo the transformation simultaneously in which case there is no phase co-existence. Thus, we believe that the coexistence between the two phases happens only in the case of non-uniform strain. It should be noted that for non-uniform strain, fractional steps in conductance could be expected. We explain below why we do not see them in our measurements.

As seen from the data presented in~\cite{soumyanarayanan2013quantum}, in on-substrate devices the domains are of the order of ten nm and more importantly, are static in time. This nanoscale phase separation is detectable in STM tunneling spectroscopic measurements which is a local probe. Our transport measurements, on the other hand, were time-dependent and performed between electrical probes separated by hundreds of nanometers. That is why we do not observe any conductance jumps from the static nanoscale phase separation seen in the on-substrate devices of the type studied in Ref.~\cite{soumyanarayanan2013quantum}. 

We note that for Co-doped samples one can expect fractional jumps. Unfortunately, we do not observe any RTN in the case of suspended Co-doped \nbse devices. We believe that this can be due to disorder inhibiting the formation of long range order in the system. It is also possible that Co doping might modify the phonon dispersion spectrum and suppress the formation of one of the two CDW phases. Further experimental and theoretical work is required to settle this issue.

\section{Quantization of conductance fluctuation:}
{We have measured different suspended devices with different thickness and observed that the two level conductance fluctuation is always present with the conductance jump of integer multiple of $e^2/h$. We have found that for thicker samples the jump is larger than thinner ones. In fig. \ref{fig:histAll} the distribution function of conductance fluctuation is plotted. The conductance jump is 1$G_0$ and 3$G_0$ for the two thin samples, S1 and S4 and is large $\sim$370$G_0$ for the thick sample S6. It can be seen from the data that the quantization is seen most clearly in the thinnest flakes. This is because, for thicker devices, the ratio of the magnitude of $1/f$ noise to the amplitude of RTN jumps is much larger than that in thin flakes. As discussed in the manuscript, this ratio is parametrized by the quantity $A/B$ (see discussion following Eqn. 1 in the main text). For example, for the sample S4 this ratio was ten times higher than that in sample S1 (Fig.~\ref{fig:AbyB}).  This higher 1/f noise cause the peaks to broaden for thicker samples.}

\section{Number of layers participating in RTN} 
{An important question is whether all the layers in the flake contribute to the observed RTN.  In case of suspended devices where RTN is seen with strain, we can envisage two possible scenarios which are as follows.  First probable case is that the bottom layer gets strained and the top layers slip on this layer to relax the strain. This will entail an energy cost, $E_{NC}$ due to non-conformity between the layers. We estimate this energy cost for relative displacement between two layers of \nbse to be  about $6.6 \times 10^{-5}$ eV/unit cell~\cite{levita2014sliding, nagapriya2008torsional,shmeliov} [See Fig.~\ref{fig:sliding}]. An alternate scenario is where all the layers get strained equally. We estimate the elastic energy cost in this process, for small strains of the magnitude applied by us (0.1\%), to be about $3\times 10^{-7}$ eV/unit cell~\cite{PhysRevB.82.155432}. It thus appears that it is energetically favourable for all the layers to strain together by the same amount. The reality of course could lie somewhere in between these two extremes - especially for very thick flakes where it is highly plausible that the strain relaxes beyond the first few layers. }

\section{Details of DFT calculations}
Electronic structure of bilayer \nbse was calculated by using density functional theory (DFT) with the generalized gradient approximation (GGA) in the parametrization of Perdew, Burke and Ernzerhof~\cite{PBE} as implemented in the Vienna ab-initio simulation package~\cite{vasp}. Projected augmented-wave (PAW) ~\cite{paw} pseudopotentials are used to describe core electrons.The electronic wavefunction is expanded using plane waves up to a cut-off energy of 600 eV. Brillouin zone sampling is done by using a 12$\times$ 12$\times$ 1 Monkhorst-Pack k-grid for the primitive unit-cell's calculations. The conjugate gradient method is used to obtain relaxed geometries. Both atomic positions and cell parameters are allowed to relax, until the forces on each atom are less than 0.01 eV/Angstrom.

Force constants were calculated for a 3$\times$3$\times$1 supercell within the framework density functional perturbation theory ~\cite{dfpt} using the VASP code. Subsequently, phonon dispersions were calculated using phonopy package[\cite{phonopy}].

We calculate the conductivity using the standard Kubo formula.
\begin{eqnarray}
	\sigma=\frac{e^{2}}{3\hbar^{2}} \int\frac{d\varepsilon}{2\pi} \left(-\frac{df(\varepsilon)}{d\varepsilon}\right)\sum_{\bf k}v_{\bf k}^{2} {\rm Tr}\big[A^{2}({\bf k},&\varepsilon)\big],
\end{eqnarray}
where $f(\epsilon)$ is the Fermi-distribution function, $v_k$ is the band velocity. $e$, and $\hbar$ have the usual meanings. $A$ is the spectral function which is obtained from the imaginary part of the Green's function obtained from Eq. 2. We averaged the conductivity over the entire Basel plane as $v_{\bf k}^2=v_{{\bf k}_x}^2+v_{{\bf k}_y}^2$. The temperature dependence of the conductivity comes from the Fermi function $f$, as well as from the $T$-dependence of the gap, and its behavior is dominated by the latter function.

\bibliography{NbSe2}

\cleardoublepage
\renewcommand{\thefigure}{S\arabic{figure}}
\setcounter{figure}{0}
\begin{figure*}
	\begin{center}
		\includegraphics[width=0.75\textwidth]{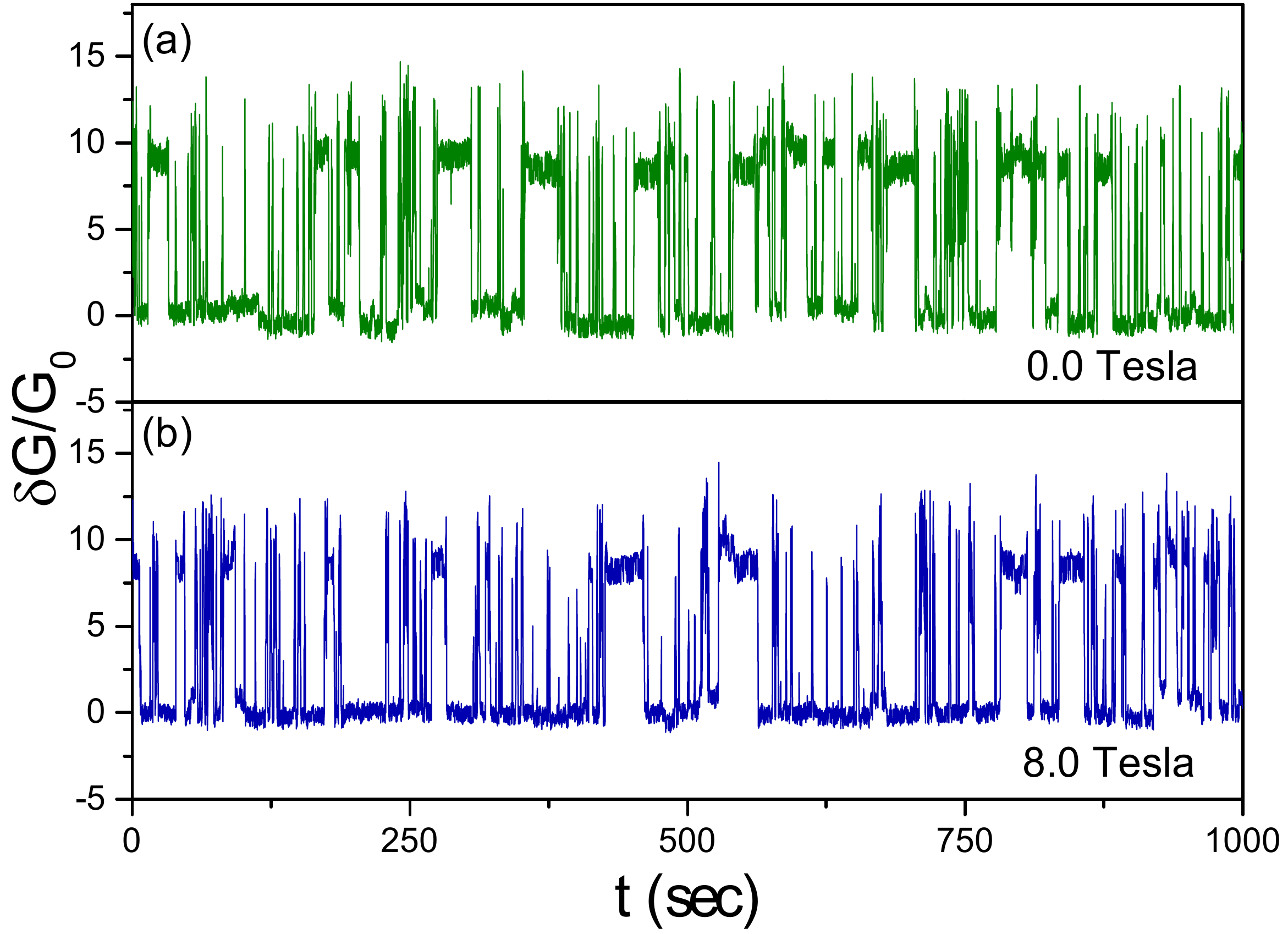}
		\small{\caption{(a) and (b) are conductance fluctuation plots measured for a clean suspended 5 layer \nbse device (S4) in presence of 0~T and 8~T perpendicular magnetic fields, respectively.
				\label{fig:S1}}}
	\end{center}
\end{figure*}

\begin{figure*}
	\begin{center}
		\includegraphics[width=0.75\textwidth]{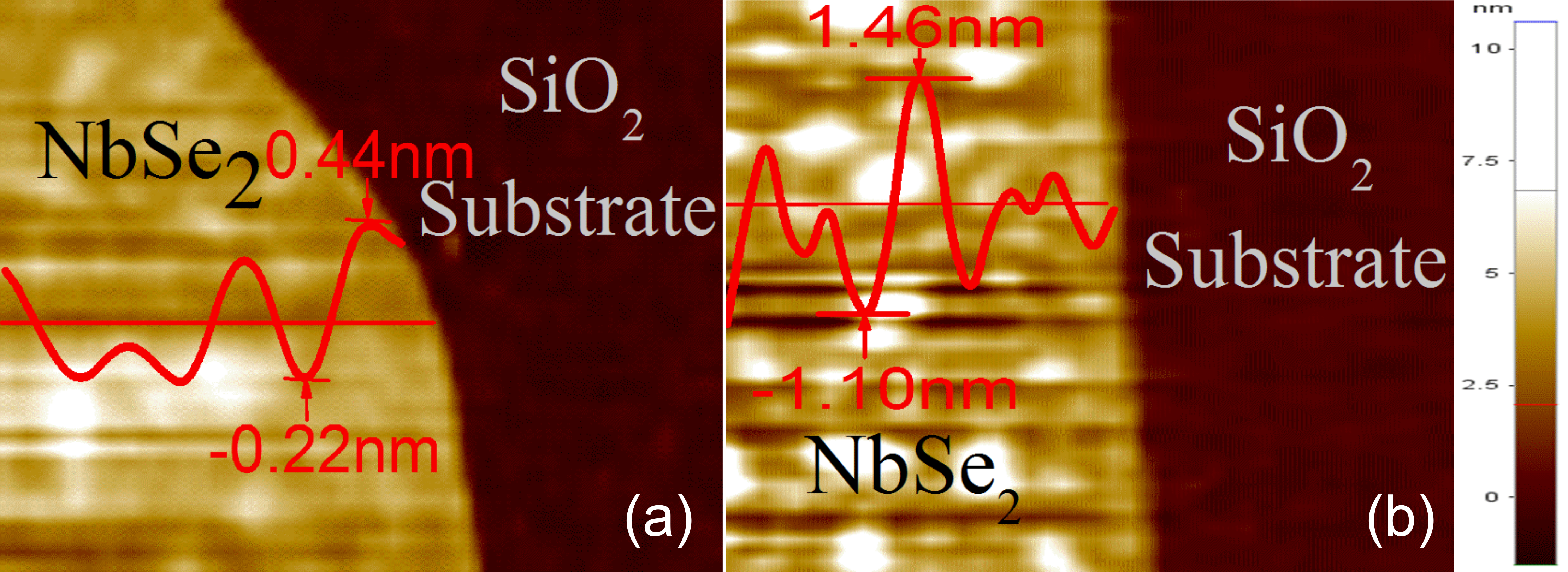}
		\small{\caption{(a) AFM topography of a flake exfoliated from bulk \nbse crystal having high superconducting $T_C$, and (b) AFM topography of a flake exfoliated from bulk \nbse crystal having low superconducting $T_C$. The red line through each image shows the line scan used to extract the rms surface roughness of the two flakes: the rms roughness of the high $T_C$ device ($\sim$ 0.44~nm) was three times lesser than that of the low $T_C$ device ($\sim$ 1.46~nm).
				\label{fig:AFM}}}
	\end{center}
\end{figure*}

\begin{figure*}
	\begin{center}
		\includegraphics[width=0.8\textwidth]{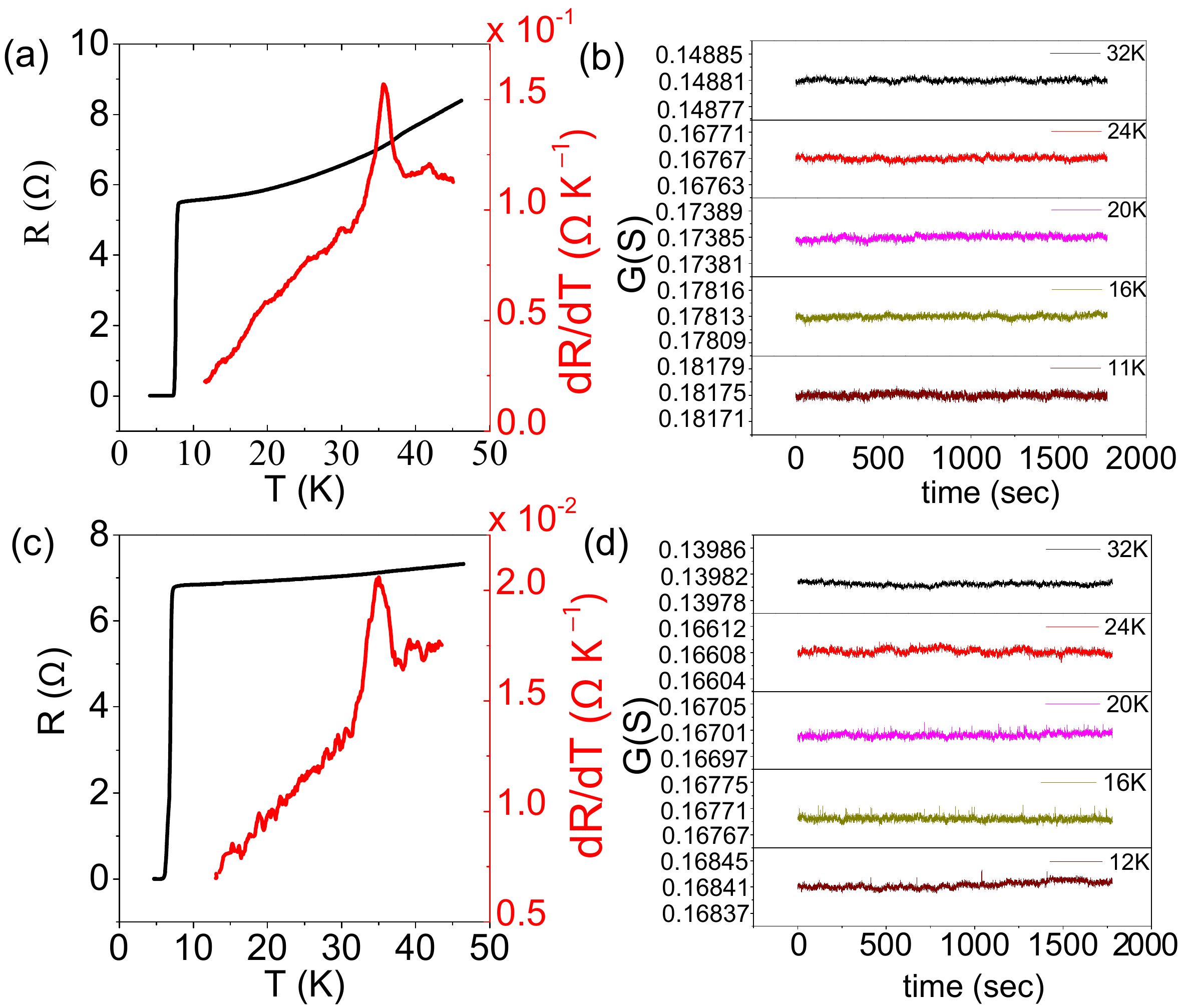}
		\small{\caption{$R$ versus $T$ plot obtained for (a) an on-substrate device  S2, and (c) a typical disordered device, S3. The $dR/dT$ plots in both cases indicate that the T$_{CDW}$ ($\sim$~35~K) remains the same as for clean suspended devices. Plots of conductance fluctuations versus time at different $T$ for (b)  on-substrate device S2 and (d) disordered device S3. In both cases, no signatures of RTN were seen. 
				\label{fig:Substrated_RT_TS}}}
	\end{center}
\end{figure*}

\begin{figure*}
	\begin{center}
		\includegraphics[width=0.8\textwidth]{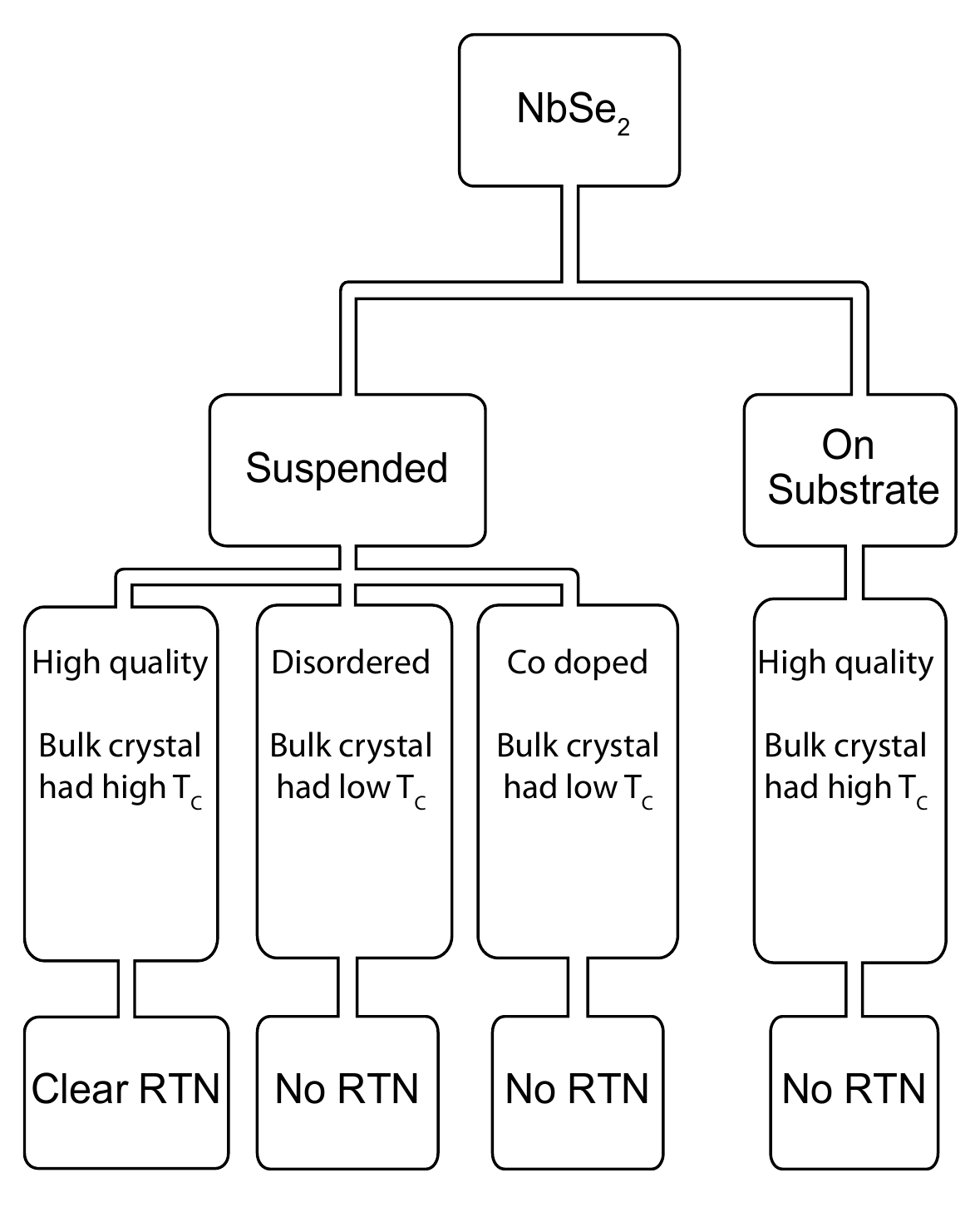}
		\small{\caption{Summary of different classes of samples measured, emphasizing that RTN was only observed in clean suspended 2H-NbSe$_2$ devices
				\label{fig:flowchart}}}
	\end{center}
\end{figure*}

\begin{figure*}[!h]
	\begin{center}
\includegraphics[width=0.75\textwidth]{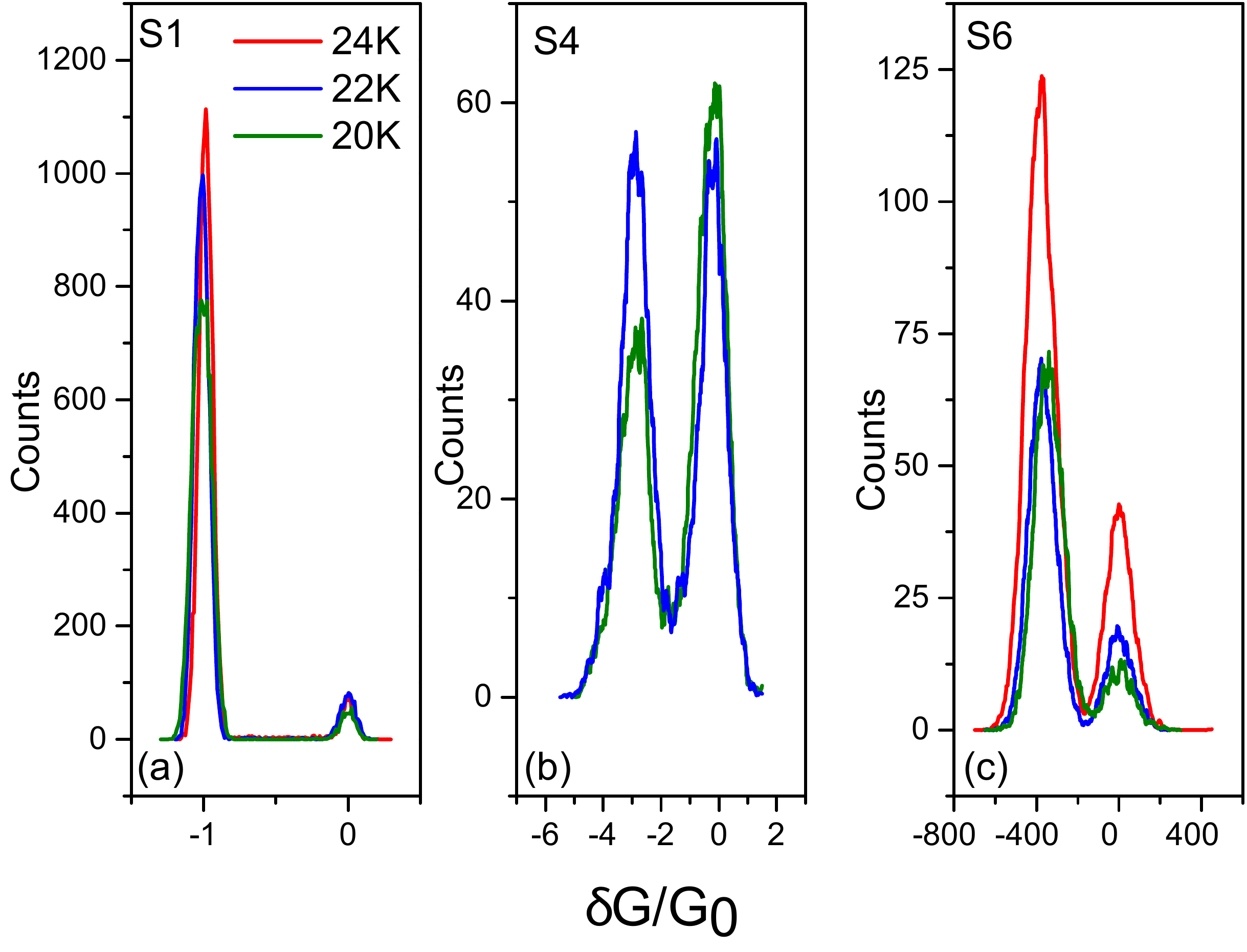}
		\small{\caption{Probability distribution function of conductance fluctuation measured for different samples. The difference of peak positions for all the samples at all the temperatures where RT is present are always integer multiple of $G_0$. The conductance jumps for different devices are (a)  $\Delta G = G_0$ for S1 (tri-layer device), (b) $\Delta G = 3G_0$ for S4 (five layer device) and, (c) $\Delta G = 370G_0$ for S6 (approximately 50~nm thick device).
				\label{fig:histAll}}}
	\end{center}
\end{figure*}

\begin{figure*}[!h]
	\begin{center}
		\includegraphics[width=0.75\textwidth]{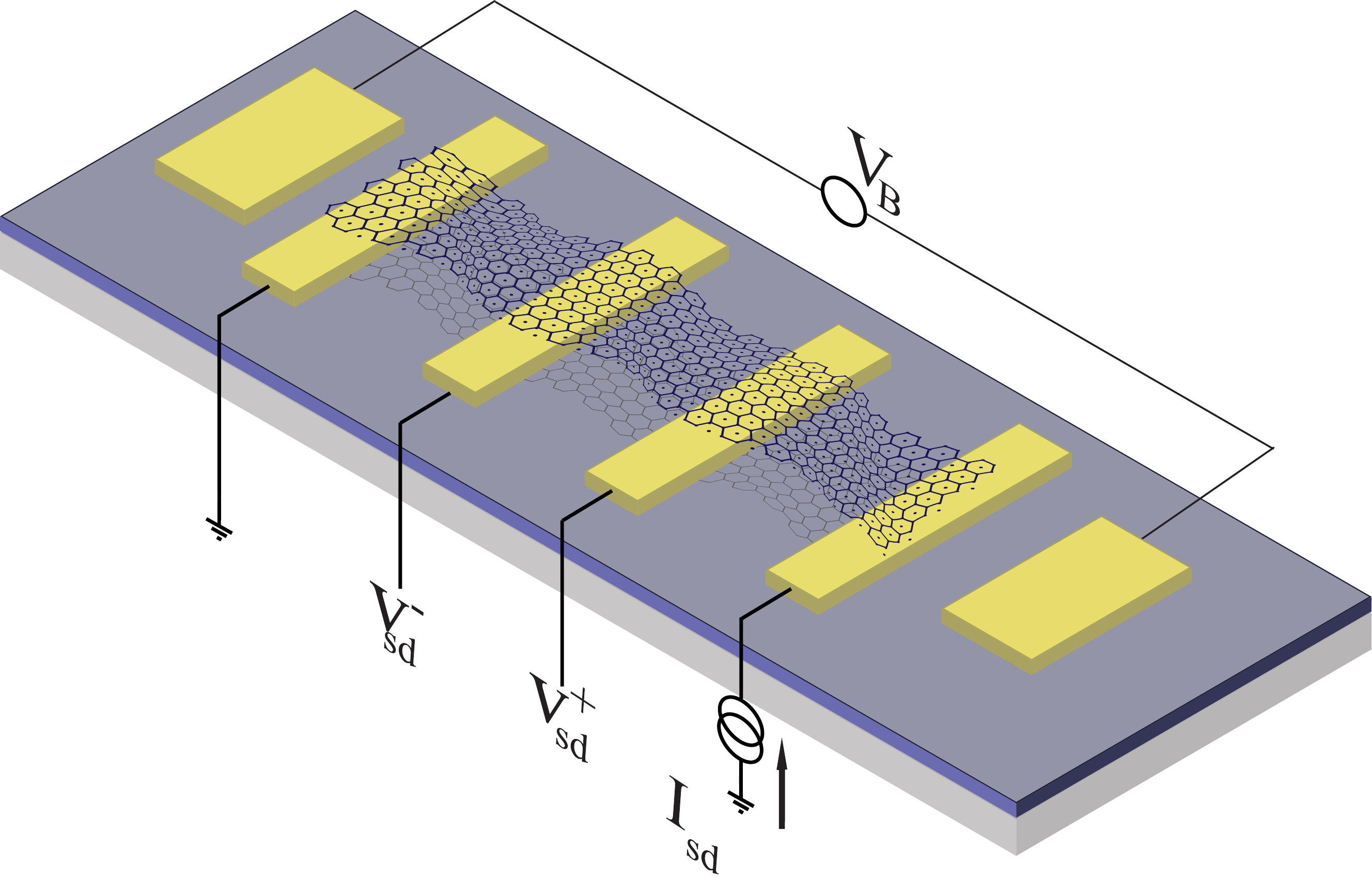}
		\small{\caption{Schematic of the ultra-thin suspended \nbse device fabricated on BTO substrate. A dc voltage $V_B$ applied across the BTO substrate was used to control the lateral strain across the device.
				\label{fig:device}}}
	\end{center}
\end{figure*}

\begin{figure*}[!h]
	\begin{center}
		\includegraphics[width=0.65\textwidth]{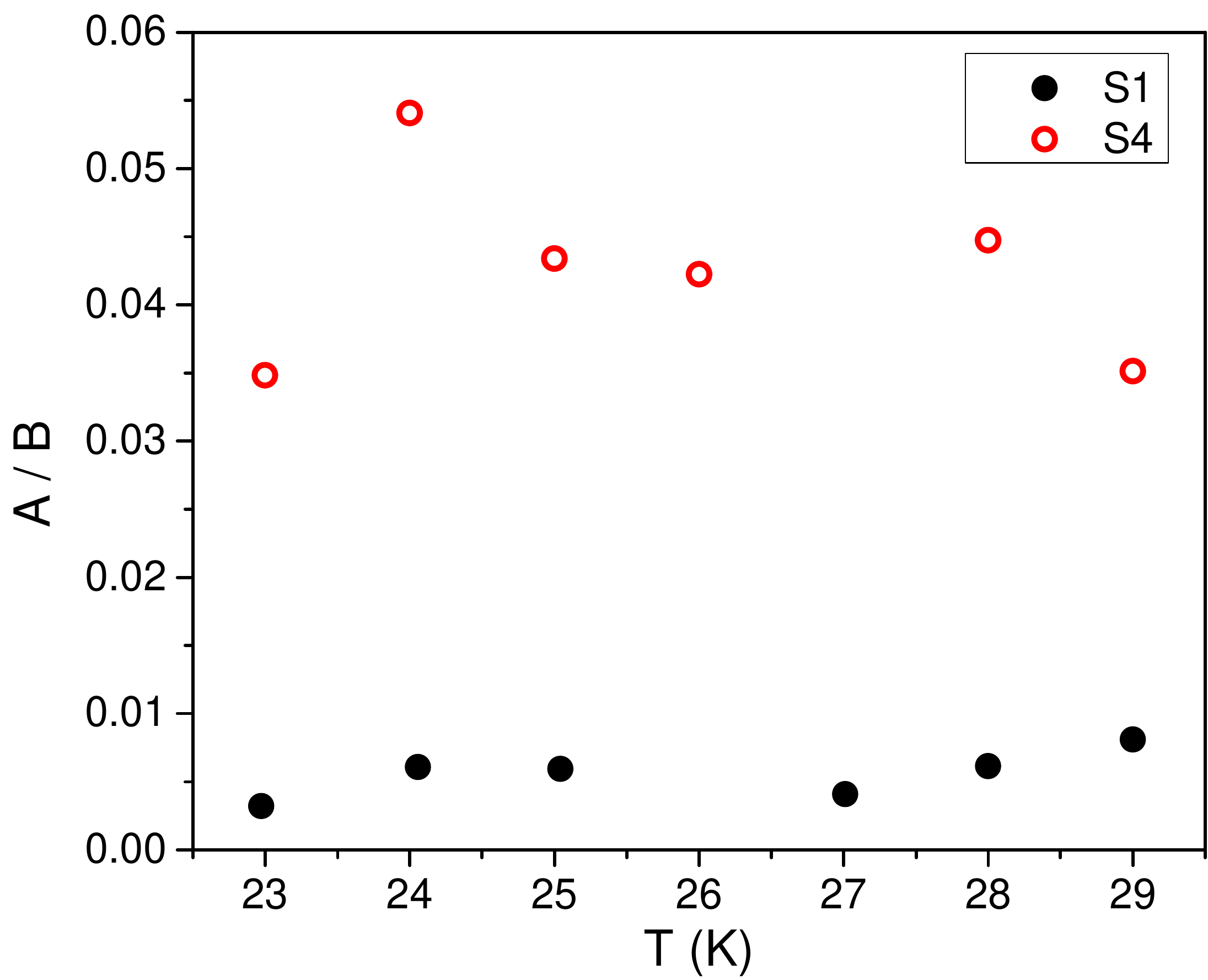}
		\small{\caption{Plot of the values of the parameter $A/B$ versus temperature extracted from noise measurements. The red open circles are for sample S4 (five layer device) while the black filled circles are for sample S1 (tri-layer device). 	\label{fig:AbyB}}}
	\end{center}
\end{figure*}

\begin{figure*}[!h]
	\begin{center}
		\includegraphics[width=0.65\textwidth]{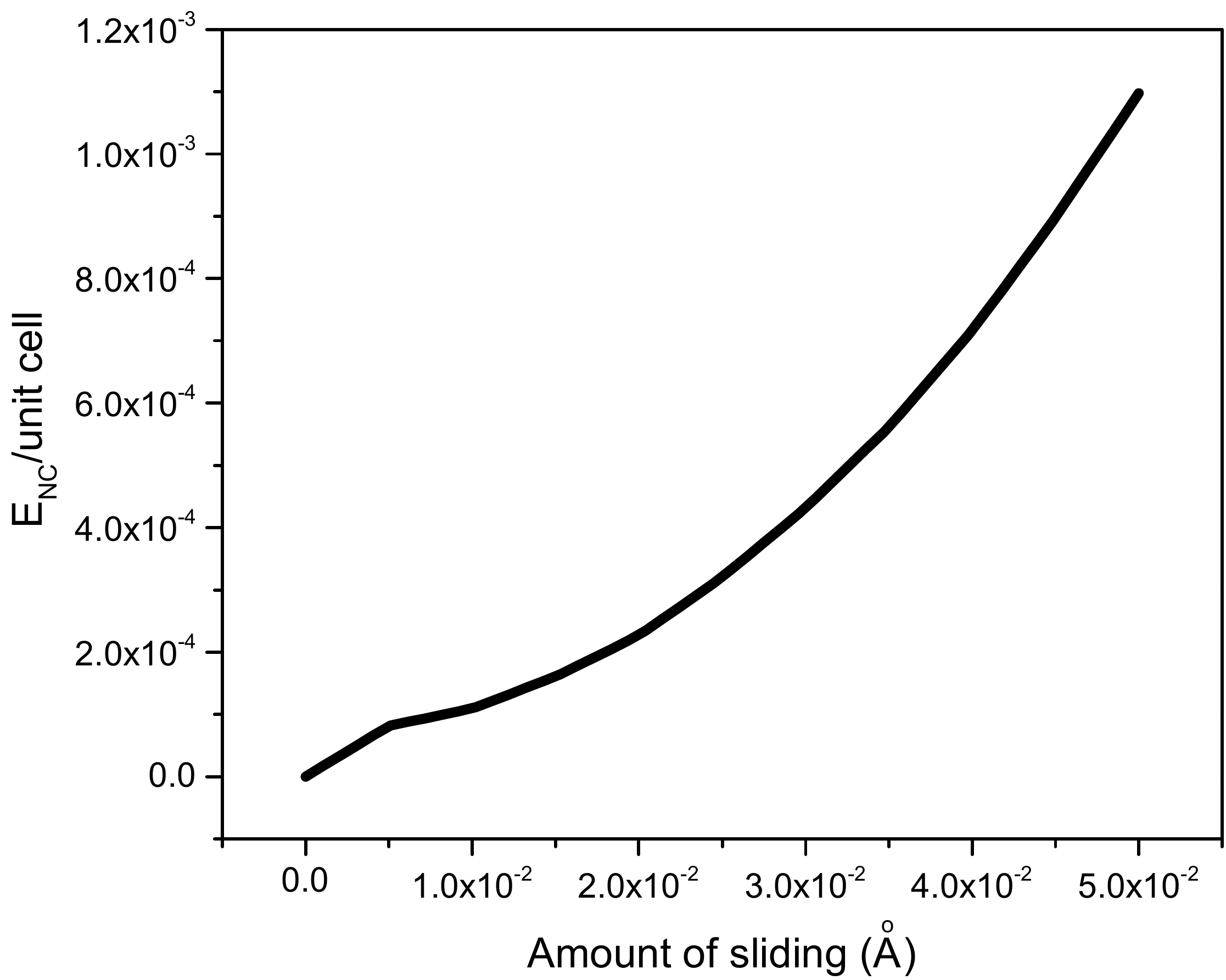}
		\small{\caption{Plot showing the dependence of $E_{NC}$ on the relative displacement between two succesive layers in 2H-NbSe$_2$. 				\label{fig:sliding}}}
	\end{center}
\end{figure*}

\bibliography{NbSe2}

\begin{thebibliography}{38}%
\makeatletter
\providecommand \@ifxundefined [1]{%
 \@ifx{#1\undefined}
}%
\providecommand \@ifnum [1]{%
 \ifnum #1\expandafter \@firstoftwo
 \else \expandafter \@secondoftwo
 \fi
}%
\providecommand \@ifx [1]{%
 \ifx #1\expandafter \@firstoftwo
 \else \expandafter \@secondoftwo
 \fi
}%
\providecommand \natexlab [1]{#1}%
\providecommand \enquote  [1]{``#1''}%
\providecommand \bibnamefont  [1]{#1}%
\providecommand \bibfnamefont [1]{#1}%
\providecommand \citenamefont [1]{#1}%
\providecommand \href@noop [0]{\@secondoftwo}%
\providecommand \href [0]{\begingroup \@sanitize@url \@href}%
\providecommand \@href[1]{\@@startlink{#1}\@@href}%
\providecommand \@@href[1]{\endgroup#1\@@endlink}%
\providecommand \@sanitize@url [0]{\catcode `\\12\catcode `\$12\catcode
  `\&12\catcode `\#12\catcode `\^12\catcode `\_12\catcode `\%12\relax}%
\providecommand \@@startlink[1]{}%
\providecommand \@@endlink[0]{}%
\providecommand \url  [0]{\begingroup\@sanitize@url \@url }%
\providecommand \@url [1]{\endgroup\@href {#1}{\urlprefix }}%
\providecommand \urlprefix  [0]{URL }%
\providecommand \Eprint [0]{\href }%
\providecommand \doibase [0]{http://dx.doi.org/}%
\providecommand \selectlanguage [0]{\@gobble}%
\providecommand \bibinfo  [0]{\@secondoftwo}%
\providecommand \bibfield  [0]{\@secondoftwo}%
\providecommand \translation [1]{[#1]}%
\providecommand \BibitemOpen [0]{}%
\providecommand \bibitemStop [0]{}%
\providecommand \bibitemNoStop [0]{.\EOS\space}%
\providecommand \EOS [0]{\spacefactor3000\relax}%
\providecommand \BibitemShut  [1]{\csname bibitem#1\endcsname}%
\let\auto@bib@innerbib\@empty
\bibitem [{\citenamefont {Gr\"uner}(1988)}]{RevModPhysGruner}%
  \BibitemOpen
  \bibfield  {author} {\bibinfo {author} {\bibfnamefont {G.}~\bibnamefont
  {Gr\"uner}},\ }\href {\doibase 10.1103/RevModPhys.60.1129} {\bibfield
  {journal} {\bibinfo  {journal} {Rev. Mod. Phys.}\ }\textbf {\bibinfo {volume}
  {60}},\ \bibinfo {pages} {1129} (\bibinfo {year} {1988})}\BibitemShut
  {NoStop}%
\bibitem [{\citenamefont {Miller}\ \emph {et~al.}(2012)\citenamefont {Miller},
  \citenamefont {Wijesinghe}, \citenamefont {Tang},\ and\ \citenamefont
  {Guloy}}]{PhysRevLett.108.036404}%
  \BibitemOpen
  \bibfield  {author} {\bibinfo {author} {\bibfnamefont {J.~H.}\ \bibnamefont
  {Miller}}, \bibinfo {author} {\bibfnamefont {A.~I.}\ \bibnamefont
  {Wijesinghe}}, \bibinfo {author} {\bibfnamefont {Z.}~\bibnamefont {Tang}}, \
  and\ \bibinfo {author} {\bibfnamefont {A.~M.}\ \bibnamefont {Guloy}},\ }\href
  {\doibase 10.1103/PhysRevLett.108.036404} {\bibfield  {journal} {\bibinfo
  {journal} {Phys. Rev. Lett.}\ }\textbf {\bibinfo {volume} {108}},\ \bibinfo
  {pages} {036404} (\bibinfo {year} {2012})}\BibitemShut {NoStop}%
\bibitem [{\citenamefont {Monceau}(2012)}]{doi:10.1080/00018732.2012.719674}%
  \BibitemOpen
  \bibfield  {author} {\bibinfo {author} {\bibfnamefont {P.}~\bibnamefont
  {Monceau}},\ }\href {\doibase 10.1080/00018732.2012.719674} {\bibfield
  {journal} {\bibinfo  {journal} {Advances in Physics}\ }\textbf {\bibinfo
  {volume} {61}},\ \bibinfo {pages} {325} (\bibinfo {year} {2012})},\ \Eprint
  {http://arxiv.org/abs/http://dx.doi.org/10.1080/00018732.2012.719674}
  {http://dx.doi.org/10.1080/00018732.2012.719674} \BibitemShut {NoStop}%
\bibitem [{\citenamefont {Peierls}(1991)}]{peierls1991more}%
  \BibitemOpen
  \bibfield  {author} {\bibinfo {author} {\bibfnamefont {R.~E.}\ \bibnamefont
  {Peierls}},\ }\href@noop {} {\emph {\bibinfo {title} {More surprises in
  theoretical physics}}},\ Vol.~\bibinfo {volume} {19}\ (\bibinfo  {publisher}
  {Princeton University Press},\ \bibinfo {year} {1991})\BibitemShut {NoStop}%
\bibitem [{\citenamefont {Matsuno}\ \emph {et~al.}(2001)\citenamefont
  {Matsuno}, \citenamefont {Fujimori}, \citenamefont {Mattheiss}, \citenamefont
  {Endoh},\ and\ \citenamefont {Nagata}}]{PhysRevB.64.115116}%
  \BibitemOpen
  \bibfield  {author} {\bibinfo {author} {\bibfnamefont {J.}~\bibnamefont
  {Matsuno}}, \bibinfo {author} {\bibfnamefont {A.}~\bibnamefont {Fujimori}},
  \bibinfo {author} {\bibfnamefont {L.~F.}\ \bibnamefont {Mattheiss}}, \bibinfo
  {author} {\bibfnamefont {R.}~\bibnamefont {Endoh}}, \ and\ \bibinfo {author}
  {\bibfnamefont {S.}~\bibnamefont {Nagata}},\ }\href {\doibase
  10.1103/PhysRevB.64.115116} {\bibfield  {journal} {\bibinfo  {journal} {Phys.
  Rev. B}\ }\textbf {\bibinfo {volume} {64}},\ \bibinfo {pages} {115116}
  (\bibinfo {year} {2001})}\BibitemShut {NoStop}%
\bibitem [{\citenamefont {Yan}\ \emph {et~al.}(2017)\citenamefont {Yan},
  \citenamefont {Iaia}, \citenamefont {Morosan}, \citenamefont {Fradkin},
  \citenamefont {Abbamonte},\ and\ \citenamefont
  {Madhavan}}]{PhysRevLett.118.106405}%
  \BibitemOpen
  \bibfield  {author} {\bibinfo {author} {\bibfnamefont {S.}~\bibnamefont
  {Yan}}, \bibinfo {author} {\bibfnamefont {D.}~\bibnamefont {Iaia}}, \bibinfo
  {author} {\bibfnamefont {E.}~\bibnamefont {Morosan}}, \bibinfo {author}
  {\bibfnamefont {E.}~\bibnamefont {Fradkin}}, \bibinfo {author} {\bibfnamefont
  {P.}~\bibnamefont {Abbamonte}}, \ and\ \bibinfo {author} {\bibfnamefont
  {V.}~\bibnamefont {Madhavan}},\ }\href {\doibase
  10.1103/PhysRevLett.118.106405} {\bibfield  {journal} {\bibinfo  {journal}
  {Phys. Rev. Lett.}\ }\textbf {\bibinfo {volume} {118}},\ \bibinfo {pages}
  {106405} (\bibinfo {year} {2017})}\BibitemShut {NoStop}%
\bibitem [{\citenamefont {Johannes}\ and\ \citenamefont
  {Mazin}(2008)}]{PhysRevB.77.165135}%
  \BibitemOpen
  \bibfield  {author} {\bibinfo {author} {\bibfnamefont {M.~D.}\ \bibnamefont
  {Johannes}}\ and\ \bibinfo {author} {\bibfnamefont {I.~I.}\ \bibnamefont
  {Mazin}},\ }\href {\doibase 10.1103/PhysRevB.77.165135} {\bibfield  {journal}
  {\bibinfo  {journal} {Phys. Rev. B}\ }\textbf {\bibinfo {volume} {77}},\
  \bibinfo {pages} {165135} (\bibinfo {year} {2008})}\BibitemShut {NoStop}%
\bibitem [{\citenamefont {Arguello}\ \emph {et~al.}(2015)\citenamefont
  {Arguello}, \citenamefont {Rosenthal}, \citenamefont {Andrade}, \citenamefont
  {Jin}, \citenamefont {Yeh}, \citenamefont {Zaki}, \citenamefont {Jia},
  \citenamefont {Cava}, \citenamefont {Fernandes}, \citenamefont {Millis},
  \citenamefont {Valla}, \citenamefont {Osgood},\ and\ \citenamefont
  {Pasupathy}}]{PhysRevLett.114.037001}%
  \BibitemOpen
  \bibfield  {author} {\bibinfo {author} {\bibfnamefont {C.~J.}\ \bibnamefont
  {Arguello}}, \bibinfo {author} {\bibfnamefont {E.~P.}\ \bibnamefont
  {Rosenthal}}, \bibinfo {author} {\bibfnamefont {E.~F.}\ \bibnamefont
  {Andrade}}, \bibinfo {author} {\bibfnamefont {W.}~\bibnamefont {Jin}},
  \bibinfo {author} {\bibfnamefont {P.~C.}\ \bibnamefont {Yeh}}, \bibinfo
  {author} {\bibfnamefont {N.}~\bibnamefont {Zaki}}, \bibinfo {author}
  {\bibfnamefont {S.}~\bibnamefont {Jia}}, \bibinfo {author} {\bibfnamefont
  {R.~J.}\ \bibnamefont {Cava}}, \bibinfo {author} {\bibfnamefont {R.~M.}\
  \bibnamefont {Fernandes}}, \bibinfo {author} {\bibfnamefont {A.~J.}\
  \bibnamefont {Millis}}, \bibinfo {author} {\bibfnamefont {T.}~\bibnamefont
  {Valla}}, \bibinfo {author} {\bibfnamefont {R.~M.}\ \bibnamefont {Osgood}}, \
  and\ \bibinfo {author} {\bibfnamefont {A.~N.}\ \bibnamefont {Pasupathy}},\
  }\href {\doibase 10.1103/PhysRevLett.114.037001} {\bibfield  {journal}
  {\bibinfo  {journal} {Phys. Rev. Lett.}\ }\textbf {\bibinfo {volume} {114}},\
  \bibinfo {pages} {037001} (\bibinfo {year} {2015})}\BibitemShut {NoStop}%
\bibitem [{\citenamefont {Valla}\ \emph {et~al.}(2004)\citenamefont {Valla},
  \citenamefont {Fedorov}, \citenamefont {Johnson}, \citenamefont {Glans},
  \citenamefont {McGuinness}, \citenamefont {Smith}, \citenamefont {Andrei},\
  and\ \citenamefont {Berger}}]{PhysRevLett.92.086401}%
  \BibitemOpen
  \bibfield  {author} {\bibinfo {author} {\bibfnamefont {T.}~\bibnamefont
  {Valla}}, \bibinfo {author} {\bibfnamefont {A.~V.}\ \bibnamefont {Fedorov}},
  \bibinfo {author} {\bibfnamefont {P.~D.}\ \bibnamefont {Johnson}}, \bibinfo
  {author} {\bibfnamefont {P.-A.}\ \bibnamefont {Glans}}, \bibinfo {author}
  {\bibfnamefont {C.}~\bibnamefont {McGuinness}}, \bibinfo {author}
  {\bibfnamefont {K.~E.}\ \bibnamefont {Smith}}, \bibinfo {author}
  {\bibfnamefont {E.~Y.}\ \bibnamefont {Andrei}}, \ and\ \bibinfo {author}
  {\bibfnamefont {H.}~\bibnamefont {Berger}},\ }\href {\doibase
  10.1103/PhysRevLett.92.086401} {\bibfield  {journal} {\bibinfo  {journal}
  {Phys. Rev. Lett.}\ }\textbf {\bibinfo {volume} {92}},\ \bibinfo {pages}
  {086401} (\bibinfo {year} {2004})}\BibitemShut {NoStop}%
\bibitem [{\citenamefont {Suderow}\ \emph {et~al.}(2005)\citenamefont
  {Suderow}, \citenamefont {Tissen}, \citenamefont {Brison}, \citenamefont
  {Mart\'{\i}nez},\ and\ \citenamefont {Vieira}}]{PhysRevLett.95.117006}%
  \BibitemOpen
  \bibfield  {author} {\bibinfo {author} {\bibfnamefont {H.}~\bibnamefont
  {Suderow}}, \bibinfo {author} {\bibfnamefont {V.~G.}\ \bibnamefont {Tissen}},
  \bibinfo {author} {\bibfnamefont {J.~P.}\ \bibnamefont {Brison}}, \bibinfo
  {author} {\bibfnamefont {J.~L.}\ \bibnamefont {Mart\'{\i}nez}}, \ and\
  \bibinfo {author} {\bibfnamefont {S.}~\bibnamefont {Vieira}},\ }\href
  {\doibase 10.1103/PhysRevLett.95.117006} {\bibfield  {journal} {\bibinfo
  {journal} {Phys. Rev. Lett.}\ }\textbf {\bibinfo {volume} {95}},\ \bibinfo
  {pages} {117006} (\bibinfo {year} {2005})}\BibitemShut {NoStop}%
\bibitem [{\citenamefont {Castro~Neto}(2001)}]{PhysRevLett.86.4382}%
  \BibitemOpen
  \bibfield  {author} {\bibinfo {author} {\bibfnamefont {A.~H.}\ \bibnamefont
  {Castro~Neto}},\ }\href {\doibase 10.1103/PhysRevLett.86.4382} {\bibfield
  {journal} {\bibinfo  {journal} {Phys. Rev. Lett.}\ }\textbf {\bibinfo
  {volume} {86}},\ \bibinfo {pages} {4382} (\bibinfo {year}
  {2001})}\BibitemShut {NoStop}%
\bibitem [{\citenamefont {Cercellier}\ \emph {et~al.}(2007)\citenamefont
  {Cercellier}, \citenamefont {Monney}, \citenamefont {Clerc}, \citenamefont
  {Battaglia}, \citenamefont {Despont}, \citenamefont {Garnier}, \citenamefont
  {Beck}, \citenamefont {Aebi}, \citenamefont {Patthey}, \citenamefont
  {Berger},\ and\ \citenamefont {Forr\'o}}]{PhysRevLett.99.146403}%
  \BibitemOpen
  \bibfield  {author} {\bibinfo {author} {\bibfnamefont {H.}~\bibnamefont
  {Cercellier}}, \bibinfo {author} {\bibfnamefont {C.}~\bibnamefont {Monney}},
  \bibinfo {author} {\bibfnamefont {F.}~\bibnamefont {Clerc}}, \bibinfo
  {author} {\bibfnamefont {C.}~\bibnamefont {Battaglia}}, \bibinfo {author}
  {\bibfnamefont {L.}~\bibnamefont {Despont}}, \bibinfo {author} {\bibfnamefont
  {M.~G.}\ \bibnamefont {Garnier}}, \bibinfo {author} {\bibfnamefont
  {H.}~\bibnamefont {Beck}}, \bibinfo {author} {\bibfnamefont {P.}~\bibnamefont
  {Aebi}}, \bibinfo {author} {\bibfnamefont {L.}~\bibnamefont {Patthey}},
  \bibinfo {author} {\bibfnamefont {H.}~\bibnamefont {Berger}}, \ and\ \bibinfo
  {author} {\bibfnamefont {L.}~\bibnamefont {Forr\'o}},\ }\href {\doibase
  10.1103/PhysRevLett.99.146403} {\bibfield  {journal} {\bibinfo  {journal}
  {Phys. Rev. Lett.}\ }\textbf {\bibinfo {volume} {99}},\ \bibinfo {pages}
  {146403} (\bibinfo {year} {2007})}\BibitemShut {NoStop}%
\bibitem [{\citenamefont {Castellan}\ \emph {et~al.}(2013)\citenamefont
  {Castellan}, \citenamefont {Rosenkranz}, \citenamefont {Osborn},
  \citenamefont {Li}, \citenamefont {Gray}, \citenamefont {Luo}, \citenamefont
  {Welp}, \citenamefont {Karapetrov}, \citenamefont {Ruff},\ and\ \citenamefont
  {van Wezel}}]{PhysRevLett.110.196404}%
  \BibitemOpen
  \bibfield  {author} {\bibinfo {author} {\bibfnamefont {J.-P.}\ \bibnamefont
  {Castellan}}, \bibinfo {author} {\bibfnamefont {S.}~\bibnamefont
  {Rosenkranz}}, \bibinfo {author} {\bibfnamefont {R.}~\bibnamefont {Osborn}},
  \bibinfo {author} {\bibfnamefont {Q.}~\bibnamefont {Li}}, \bibinfo {author}
  {\bibfnamefont {K.~E.}\ \bibnamefont {Gray}}, \bibinfo {author}
  {\bibfnamefont {X.}~\bibnamefont {Luo}}, \bibinfo {author} {\bibfnamefont
  {U.}~\bibnamefont {Welp}}, \bibinfo {author} {\bibfnamefont {G.}~\bibnamefont
  {Karapetrov}}, \bibinfo {author} {\bibfnamefont {J.~P.~C.}\ \bibnamefont
  {Ruff}}, \ and\ \bibinfo {author} {\bibfnamefont {J.}~\bibnamefont {van
  Wezel}},\ }\href {\doibase 10.1103/PhysRevLett.110.196404} {\bibfield
  {journal} {\bibinfo  {journal} {Phys. Rev. Lett.}\ }\textbf {\bibinfo
  {volume} {110}},\ \bibinfo {pages} {196404} (\bibinfo {year}
  {2013})}\BibitemShut {NoStop}%
\bibitem [{\citenamefont {Ishioka}\ \emph {et~al.}(2010)\citenamefont
  {Ishioka}, \citenamefont {Liu}, \citenamefont {Shimatake}, \citenamefont
  {Kurosawa}, \citenamefont {Ichimura}, \citenamefont {Toda}, \citenamefont
  {Oda},\ and\ \citenamefont {Tanda}}]{PhysRevLett.105.176401}%
  \BibitemOpen
  \bibfield  {author} {\bibinfo {author} {\bibfnamefont {J.}~\bibnamefont
  {Ishioka}}, \bibinfo {author} {\bibfnamefont {Y.~H.}\ \bibnamefont {Liu}},
  \bibinfo {author} {\bibfnamefont {K.}~\bibnamefont {Shimatake}}, \bibinfo
  {author} {\bibfnamefont {T.}~\bibnamefont {Kurosawa}}, \bibinfo {author}
  {\bibfnamefont {K.}~\bibnamefont {Ichimura}}, \bibinfo {author}
  {\bibfnamefont {Y.}~\bibnamefont {Toda}}, \bibinfo {author} {\bibfnamefont
  {M.}~\bibnamefont {Oda}}, \ and\ \bibinfo {author} {\bibfnamefont
  {S.}~\bibnamefont {Tanda}},\ }\href {\doibase 10.1103/PhysRevLett.105.176401}
  {\bibfield  {journal} {\bibinfo  {journal} {Phys. Rev. Lett.}\ }\textbf
  {\bibinfo {volume} {105}},\ \bibinfo {pages} {176401} (\bibinfo {year}
  {2010})}\BibitemShut {NoStop}%
\bibitem [{\citenamefont {Ugeda}\ \emph {et~al.}(2016)\citenamefont {Ugeda},
  \citenamefont {Bradley}, \citenamefont {Zhang}, \citenamefont {Onishi},
  \citenamefont {Chen}, \citenamefont {Ruan}, \citenamefont
  {Ojeda-Aristizabal}, \citenamefont {Ryu}, \citenamefont {Edmonds},
  \citenamefont {Tsai} \emph {et~al.}}]{ugeda2016characterization}%
  \BibitemOpen
  \bibfield  {author} {\bibinfo {author} {\bibfnamefont {M.~M.}\ \bibnamefont
  {Ugeda}}, \bibinfo {author} {\bibfnamefont {A.~J.}\ \bibnamefont {Bradley}},
  \bibinfo {author} {\bibfnamefont {Y.}~\bibnamefont {Zhang}}, \bibinfo
  {author} {\bibfnamefont {S.}~\bibnamefont {Onishi}}, \bibinfo {author}
  {\bibfnamefont {Y.}~\bibnamefont {Chen}}, \bibinfo {author} {\bibfnamefont
  {W.}~\bibnamefont {Ruan}}, \bibinfo {author} {\bibfnamefont {C.}~\bibnamefont
  {Ojeda-Aristizabal}}, \bibinfo {author} {\bibfnamefont {H.}~\bibnamefont
  {Ryu}}, \bibinfo {author} {\bibfnamefont {M.~T.}\ \bibnamefont {Edmonds}},
  \bibinfo {author} {\bibfnamefont {H.-Z.}\ \bibnamefont {Tsai}},  \emph
  {et~al.},\ }\href@noop {} {\bibfield  {journal} {\bibinfo  {journal} {Nature
  Physics}\ }\textbf {\bibinfo {volume} {12}},\ \bibinfo {pages} {92} (\bibinfo
  {year} {2016})}\BibitemShut {NoStop}%
\bibitem [{\citenamefont {Weber}\ \emph {et~al.}(2013)\citenamefont {Weber},
  \citenamefont {Hott}, \citenamefont {Heid}, \citenamefont {Bohnen},
  \citenamefont {Rosenkranz}, \citenamefont {Castellan}, \citenamefont
  {Osborn}, \citenamefont {Said}, \citenamefont {Leu},\ and\ \citenamefont
  {Reznik}}]{PhysRevB.87.245111}%
  \BibitemOpen
  \bibfield  {author} {\bibinfo {author} {\bibfnamefont {F.}~\bibnamefont
  {Weber}}, \bibinfo {author} {\bibfnamefont {R.}~\bibnamefont {Hott}},
  \bibinfo {author} {\bibfnamefont {R.}~\bibnamefont {Heid}}, \bibinfo {author}
  {\bibfnamefont {K.-P.}\ \bibnamefont {Bohnen}}, \bibinfo {author}
  {\bibfnamefont {S.}~\bibnamefont {Rosenkranz}}, \bibinfo {author}
  {\bibfnamefont {J.-P.}\ \bibnamefont {Castellan}}, \bibinfo {author}
  {\bibfnamefont {R.}~\bibnamefont {Osborn}}, \bibinfo {author} {\bibfnamefont
  {A.~H.}\ \bibnamefont {Said}}, \bibinfo {author} {\bibfnamefont {B.~M.}\
  \bibnamefont {Leu}}, \ and\ \bibinfo {author} {\bibfnamefont
  {D.}~\bibnamefont {Reznik}},\ }\href {\doibase 10.1103/PhysRevB.87.245111}
  {\bibfield  {journal} {\bibinfo  {journal} {Phys. Rev. B}\ }\textbf {\bibinfo
  {volume} {87}},\ \bibinfo {pages} {245111} (\bibinfo {year}
  {2013})}\BibitemShut {NoStop}%
\bibitem [{\citenamefont {Borisenko}\ \emph {et~al.}(2009)\citenamefont
  {Borisenko}, \citenamefont {Kordyuk}, \citenamefont {Zabolotnyy},
  \citenamefont {Inosov}, \citenamefont {Evtushinsky}, \citenamefont
  {B{\"u}chner}, \citenamefont {Yaresko}, \citenamefont {Varykhalov},
  \citenamefont {Follath}, \citenamefont {Eberhardt} \emph
  {et~al.}}]{borisenko2009two}%
  \BibitemOpen
  \bibfield  {author} {\bibinfo {author} {\bibfnamefont {S.}~\bibnamefont
  {Borisenko}}, \bibinfo {author} {\bibfnamefont {A.}~\bibnamefont {Kordyuk}},
  \bibinfo {author} {\bibfnamefont {V.}~\bibnamefont {Zabolotnyy}}, \bibinfo
  {author} {\bibfnamefont {D.}~\bibnamefont {Inosov}}, \bibinfo {author}
  {\bibfnamefont {D.}~\bibnamefont {Evtushinsky}}, \bibinfo {author}
  {\bibfnamefont {B.}~\bibnamefont {B{\"u}chner}}, \bibinfo {author}
  {\bibfnamefont {A.}~\bibnamefont {Yaresko}}, \bibinfo {author} {\bibfnamefont
  {A.}~\bibnamefont {Varykhalov}}, \bibinfo {author} {\bibfnamefont
  {R.}~\bibnamefont {Follath}}, \bibinfo {author} {\bibfnamefont
  {W.}~\bibnamefont {Eberhardt}},  \emph {et~al.},\ }\href@noop {} {\bibfield
  {journal} {\bibinfo  {journal} {Physical review letters}\ }\textbf {\bibinfo
  {volume} {102}},\ \bibinfo {pages} {166402} (\bibinfo {year}
  {2009})}\BibitemShut {NoStop}%
\bibitem [{\citenamefont {Johannes}\ \emph {et~al.}(2006)\citenamefont
  {Johannes}, \citenamefont {Mazin},\ and\ \citenamefont
  {Howells}}]{johannes2006fermi}%
  \BibitemOpen
  \bibfield  {author} {\bibinfo {author} {\bibfnamefont {M.}~\bibnamefont
  {Johannes}}, \bibinfo {author} {\bibfnamefont {I.}~\bibnamefont {Mazin}}, \
  and\ \bibinfo {author} {\bibfnamefont {C.}~\bibnamefont {Howells}},\
  }\href@noop {} {\bibfield  {journal} {\bibinfo  {journal} {Physical Review
  B}\ }\textbf {\bibinfo {volume} {73}},\ \bibinfo {pages} {205102} (\bibinfo
  {year} {2006})}\BibitemShut {NoStop}%
\bibitem [{\citenamefont {Weber}\ \emph {et~al.}(2011)\citenamefont {Weber},
  \citenamefont {Rosenkranz}, \citenamefont {Castellan}, \citenamefont
  {Osborn}, \citenamefont {Hott}, \citenamefont {Heid}, \citenamefont {Bohnen},
  \citenamefont {Egami}, \citenamefont {Said},\ and\ \citenamefont
  {Reznik}}]{weber2011extended}%
  \BibitemOpen
  \bibfield  {author} {\bibinfo {author} {\bibfnamefont {F.}~\bibnamefont
  {Weber}}, \bibinfo {author} {\bibfnamefont {S.}~\bibnamefont {Rosenkranz}},
  \bibinfo {author} {\bibfnamefont {J.-P.}\ \bibnamefont {Castellan}}, \bibinfo
  {author} {\bibfnamefont {R.}~\bibnamefont {Osborn}}, \bibinfo {author}
  {\bibfnamefont {R.}~\bibnamefont {Hott}}, \bibinfo {author} {\bibfnamefont
  {R.}~\bibnamefont {Heid}}, \bibinfo {author} {\bibfnamefont {K.-P.}\
  \bibnamefont {Bohnen}}, \bibinfo {author} {\bibfnamefont {T.}~\bibnamefont
  {Egami}}, \bibinfo {author} {\bibfnamefont {A.}~\bibnamefont {Said}}, \ and\
  \bibinfo {author} {\bibfnamefont {D.}~\bibnamefont {Reznik}},\ }\href@noop {}
  {\bibfield  {journal} {\bibinfo  {journal} {Phys. Rev. lett.}\
  }\textbf {\bibinfo {volume} {107}},\ \bibinfo {pages} {107403} (\bibinfo
  {year} {2011})}\BibitemShut {NoStop}%
\bibitem [{\citenamefont {Flicker}\ and\ \citenamefont
  {Van~Wezel}(2015)}]{flicker2015charge}%
  \BibitemOpen
  \bibfield  {author} {\bibinfo {author} {\bibfnamefont {F.}~\bibnamefont
  {Flicker}}\ and\ \bibinfo {author} {\bibfnamefont {J.}~\bibnamefont
  {Van~Wezel}},\ }\href@noop {} {\bibfield  {journal} {\bibinfo  {journal}
  {Nature communications}\ }\textbf {\bibinfo {volume} {6}} (\bibinfo {year}
  {2015})}\BibitemShut {NoStop}%
\bibitem [{\citenamefont {Soumyanarayanan}\ \emph {et~al.}(2013)\citenamefont
  {Soumyanarayanan}, \citenamefont {Yee}, \citenamefont {He}, \citenamefont
  {van Wezel}, \citenamefont {Rahn}, \citenamefont {Rossnagel}, \citenamefont
  {Hudson}, \citenamefont {Norman},\ and\ \citenamefont
  {Hoffman}}]{soumyanarayanan2013quantum}%
  \BibitemOpen
  \bibfield  {author} {\bibinfo {author} {\bibfnamefont {A.}~\bibnamefont
  {Soumyanarayanan}}, \bibinfo {author} {\bibfnamefont {M.~M.}\ \bibnamefont
  {Yee}}, \bibinfo {author} {\bibfnamefont {Y.}~\bibnamefont {He}}, \bibinfo
  {author} {\bibfnamefont {J.}~\bibnamefont {van Wezel}}, \bibinfo {author}
  {\bibfnamefont {D.~J.}\ \bibnamefont {Rahn}}, \bibinfo {author}
  {\bibfnamefont {K.}~\bibnamefont {Rossnagel}}, \bibinfo {author}
  {\bibfnamefont {E.}~\bibnamefont {Hudson}}, \bibinfo {author} {\bibfnamefont
  {M.~R.}\ \bibnamefont {Norman}}, \ and\ \bibinfo {author} {\bibfnamefont
  {J.~E.}\ \bibnamefont {Hoffman}},\ }\href@noop {} {\bibfield  {journal}
  {\bibinfo  {journal} {Proceedings of the National Academy of Sciences}\
  }\textbf {\bibinfo {volume} {110}},\ \bibinfo {pages} {1623} (\bibinfo {year}
  {2013})}\BibitemShut {NoStop}%
\bibitem [{\citenamefont {Flicker}\ and\ \citenamefont {van
  Wezel}(2015)}]{flicker2015charge2}%
  \BibitemOpen
  \bibfield  {author} {\bibinfo {author} {\bibfnamefont {F.}~\bibnamefont
  {Flicker}}\ and\ \bibinfo {author} {\bibfnamefont {J.}~\bibnamefont {van
  Wezel}},\ }\href@noop {} {\bibfield  {journal} {\bibinfo  {journal} {Physical
  Review B}\ }\textbf {\bibinfo {volume} {92}},\ \bibinfo {pages} {201103}
  (\bibinfo {year} {2015})}\BibitemShut {NoStop}%
\bibitem [{\citenamefont {Flicker}\ and\ \citenamefont {van
  Wezel}(2016)}]{flicker2016charge3}%
  \BibitemOpen
  \bibfield  {author} {\bibinfo {author} {\bibfnamefont {F.}~\bibnamefont
  {Flicker}}\ and\ \bibinfo {author} {\bibfnamefont {J.}~\bibnamefont {van
  Wezel}},\ }\href@noop {} {\bibfield  {journal} {\bibinfo  {journal} {Physical
  Review B}\ }\textbf {\bibinfo {volume} {94}},\ \bibinfo {pages} {235135}
  (\bibinfo {year} {2016})}\BibitemShut {NoStop}%
\bibitem [{\citenamefont {Ghosh}\ \emph {et~al.}(2004)\citenamefont {Ghosh},
  \citenamefont {Kar}, \citenamefont {Bid},\ and\ \citenamefont
  {Raychaudhuri}}]{ghosh2004set}%
  \BibitemOpen
  \bibfield  {author} {\bibinfo {author} {\bibfnamefont {A.}~\bibnamefont
  {Ghosh}}, \bibinfo {author} {\bibfnamefont {S.}~\bibnamefont {Kar}}, \bibinfo
  {author} {\bibfnamefont {A.}~\bibnamefont {Bid}}, \ and\ \bibinfo {author}
  {\bibfnamefont {A.}~\bibnamefont {Raychaudhuri}},\ }\href@noop {} {\bibfield
  {journal} {\bibinfo  {journal} {arXiv preprint cond-mat/0402130}\ } (\bibinfo
  {year} {2004})}\BibitemShut {NoStop}%
\bibitem [{\citenamefont {Ganguli}\ \emph {et~al.}(2016)\citenamefont
  {Ganguli}, \citenamefont {Singh}, \citenamefont {Roy}, \citenamefont {Bagwe},
  \citenamefont {Bala}, \citenamefont {Thamizhavel},\ and\ \citenamefont
  {Raychaudhuri}}]{ganguli2016disorder}%
  \BibitemOpen
  \bibfield  {author} {\bibinfo {author} {\bibfnamefont {S.~C.}\ \bibnamefont
  {Ganguli}}, \bibinfo {author} {\bibfnamefont {H.}~\bibnamefont {Singh}},
  \bibinfo {author} {\bibfnamefont {I.}~\bibnamefont {Roy}}, \bibinfo {author}
  {\bibfnamefont {V.}~\bibnamefont {Bagwe}}, \bibinfo {author} {\bibfnamefont
  {D.}~\bibnamefont {Bala}}, \bibinfo {author} {\bibfnamefont {A.}~\bibnamefont
  {Thamizhavel}}, \ and\ \bibinfo {author} {\bibfnamefont {P.}~\bibnamefont
  {Raychaudhuri}},\ }\href@noop {} {\bibfield  {journal} {\bibinfo  {journal}
  {Physical Review B}\ }\textbf {\bibinfo {volume} {93}},\ \bibinfo {pages}
  {144503} (\bibinfo {year} {2016})}\BibitemShut {NoStop}%
\bibitem [{\citenamefont {Fagerquist}\ \emph {et~al.}(1989)\citenamefont
  {Fagerquist}, \citenamefont {Kirby},\ and\ \citenamefont
  {Pearlstein}}]{fagerquist1989metastable}%
  \BibitemOpen
  \bibfield  {author} {\bibinfo {author} {\bibfnamefont {R.}~\bibnamefont
  {Fagerquist}}, \bibinfo {author} {\bibfnamefont {R.~D.}\ \bibnamefont
  {Kirby}}, \ and\ \bibinfo {author} {\bibfnamefont {E.~A.}\ \bibnamefont
  {Pearlstein}},\ }\href@noop {} {\bibfield  {journal} {\bibinfo  {journal}
  {Physical Review B}\ }\textbf {\bibinfo {volume} {39}},\ \bibinfo {pages}
  {5139} (\bibinfo {year} {1989})}\BibitemShut {NoStop}%
\bibitem [{\citenamefont {Koushik}\ \emph {et~al.}(2013)\citenamefont
  {Koushik}, \citenamefont {Kumar}, \citenamefont {Amin}, \citenamefont
  {Mondal}, \citenamefont {Jesudasan}, \citenamefont {Bid}, \citenamefont
  {Raychaudhuri},\ and\ \citenamefont {Ghosh}}]{koushik2013correlated}%
  \BibitemOpen
  \bibfield  {author} {\bibinfo {author} {\bibfnamefont {R.}~\bibnamefont
  {Koushik}}, \bibinfo {author} {\bibfnamefont {S.}~\bibnamefont {Kumar}},
  \bibinfo {author} {\bibfnamefont {K.~R.}\ \bibnamefont {Amin}}, \bibinfo
  {author} {\bibfnamefont {M.}~\bibnamefont {Mondal}}, \bibinfo {author}
  {\bibfnamefont {J.}~\bibnamefont {Jesudasan}}, \bibinfo {author}
  {\bibfnamefont {A.}~\bibnamefont {Bid}}, \bibinfo {author} {\bibfnamefont
  {P.}~\bibnamefont {Raychaudhuri}}, \ and\ \bibinfo {author} {\bibfnamefont
  {A.}~\bibnamefont {Ghosh}},\ }\href@noop {} {\bibfield  {journal} {\bibinfo
  {journal} {Physical review letters}\ }\textbf {\bibinfo {volume} {111}},\
  \bibinfo {pages} {197001} (\bibinfo {year} {2013})}\BibitemShut {NoStop}%
\bibitem [{\citenamefont {Daptary}\ \emph {et~al.}(2016)\citenamefont
  {Daptary}, \citenamefont {Kumar}, \citenamefont {Kumar}, \citenamefont
  {Dogra}, \citenamefont {Mohanta}, \citenamefont {Taraphder},\ and\
  \citenamefont {Bid}}]{PhysRevB.94.085104}%
  \BibitemOpen
  \bibfield  {author} {\bibinfo {author} {\bibfnamefont {G.~N.}\ \bibnamefont
  {Daptary}}, \bibinfo {author} {\bibfnamefont {S.}~\bibnamefont {Kumar}},
  \bibinfo {author} {\bibfnamefont {P.}~\bibnamefont {Kumar}}, \bibinfo
  {author} {\bibfnamefont {A.}~\bibnamefont {Dogra}}, \bibinfo {author}
  {\bibfnamefont {N.}~\bibnamefont {Mohanta}}, \bibinfo {author} {\bibfnamefont
  {A.}~\bibnamefont {Taraphder}}, \ and\ \bibinfo {author} {\bibfnamefont
  {A.}~\bibnamefont {Bid}},\ }\href {\doibase 10.1103/PhysRevB.94.085104}
  {\bibfield  {journal} {\bibinfo  {journal} {Phys. Rev. B}\ }\textbf {\bibinfo
  {volume} {94}},\ \bibinfo {pages} {085104} (\bibinfo {year}
  {2016})}\BibitemShut {NoStop}%
\bibitem [{\citenamefont {Shi}\ \emph {et~al.}(2016)\citenamefont {Shi},
  \citenamefont {Shi},\ and\ \citenamefont {Popovi{\'c}}}]{shi2016evidence}%
  \BibitemOpen
  \bibfield  {author} {\bibinfo {author} {\bibfnamefont {Z.}~\bibnamefont
  {Shi}}, \bibinfo {author} {\bibfnamefont {X.}~\bibnamefont {Shi}}, \ and\
  \bibinfo {author} {\bibfnamefont {D.}~\bibnamefont {Popovi{\'c}}},\
  }\href@noop {} {\bibfield  {journal} {\bibinfo  {journal} {Physical Review
  B}\ }\textbf {\bibinfo {volume} {94}},\ \bibinfo {pages} {134503} (\bibinfo
  {year} {2016})}\BibitemShut {NoStop}%
\bibitem [{\citenamefont {Bloom}\ \emph
  {et~al.}(1994{\natexlab{a}})\citenamefont {Bloom}, \citenamefont {Marley},\
  and\ \citenamefont {Weissman}}]{bloom1994discrete}%
  \BibitemOpen
  \bibfield  {author} {\bibinfo {author} {\bibfnamefont {I.}~\bibnamefont
  {Bloom}}, \bibinfo {author} {\bibfnamefont {A.}~\bibnamefont {Marley}}, \
  and\ \bibinfo {author} {\bibfnamefont {M.}~\bibnamefont {Weissman}},\
  }\href@noop {} {\bibfield  {journal} {\bibinfo  {journal} {Physical Review
  B}\ }\textbf {\bibinfo {volume} {50}},\ \bibinfo {pages} {5081} (\bibinfo
  {year} {1994}{\natexlab{a}})}\BibitemShut {NoStop}%
\bibitem [{\citenamefont {Marley}\ \emph {et~al.}(1994)\citenamefont {Marley},
  \citenamefont {Bloom},\ and\ \citenamefont
  {Weissman}}]{marley1994temperature}%
  \BibitemOpen
  \bibfield  {author} {\bibinfo {author} {\bibfnamefont {A.}~\bibnamefont
  {Marley}}, \bibinfo {author} {\bibfnamefont {I.}~\bibnamefont {Bloom}}, \
  and\ \bibinfo {author} {\bibfnamefont {M.}~\bibnamefont {Weissman}},\
  }\href@noop {} {\bibfield  {journal} {\bibinfo  {journal} {Physical Review
  B}\ }\textbf {\bibinfo {volume} {49}},\ \bibinfo {pages} {16156} (\bibinfo
  {year} {1994})}\BibitemShut {NoStop}%
\bibitem [{\citenamefont {Bloom}\ \emph
  {et~al.}(1994{\natexlab{b}})\citenamefont {Bloom}, \citenamefont {Marley},\
  and\ \citenamefont {Weissman}}]{bloom1994correlation}%
  \BibitemOpen
  \bibfield  {author} {\bibinfo {author} {\bibfnamefont {I.}~\bibnamefont
  {Bloom}}, \bibinfo {author} {\bibfnamefont {A.}~\bibnamefont {Marley}}, \
  and\ \bibinfo {author} {\bibfnamefont {M.}~\bibnamefont {Weissman}},\
  }\href@noop {} {\bibfield  {journal} {\bibinfo  {journal} {Physical Review
  B}\ }\textbf {\bibinfo {volume} {50}},\ \bibinfo {pages} {12218} (\bibinfo
  {year} {1994}{\natexlab{b}})}\BibitemShut {NoStop}%
\bibitem [{\citenamefont {Za{\"\i}tsev-Zotov}\ and\ \citenamefont
  {Pokrovski{\"\i}}(1989)}]{pokrovskii1989solitary}%
  \BibitemOpen
  \bibfield  {author} {\bibinfo {author} {\bibfnamefont {S.~V.}\ \bibnamefont
  {Za{\"\i}tsev-Zotov}}\ and\ \bibinfo {author} {\bibfnamefont {V.~Y.}\
  \bibnamefont {Pokrovski{\"\i}}},\ }\href@noop {} {\bibfield  {journal}
  {\bibinfo  {journal} {JETP Lett}\ }\textbf {\bibinfo {volume} {49}} (\bibinfo
  {year} {1989})}\BibitemShut {NoStop}%
\bibitem [{\citenamefont {Kummamuru}\ and\ \citenamefont
  {Soh}(2008)}]{kummamuru2008electrical}%
  \BibitemOpen
  \bibfield  {author} {\bibinfo {author} {\bibfnamefont {R.~K.}\ \bibnamefont
  {Kummamuru}}\ and\ \bibinfo {author} {\bibfnamefont {Y.-A.}\ \bibnamefont
  {Soh}},\ }\href@noop {} {\bibfield  {journal} {\bibinfo  {journal} {Nature}\
  }\textbf {\bibinfo {volume} {452}},\ \bibinfo {pages} {859} (\bibinfo {year}
  {2008})}\BibitemShut {NoStop}%
\bibitem [{\citenamefont {Jaramillo}\ \emph {et~al.}(2007)\citenamefont
  {Jaramillo}, \citenamefont {Rosenbaum}, \citenamefont {Isaacs}, \citenamefont
  {Shpyrko}, \citenamefont {Evans}, \citenamefont {Aeppli},\ and\ \citenamefont
  {Cai}}]{PhysRevLett.98.117206}%
  \BibitemOpen
  \bibfield  {author} {\bibinfo {author} {\bibfnamefont {R.}~\bibnamefont
  {Jaramillo}}, \bibinfo {author} {\bibfnamefont {T.~F.}\ \bibnamefont
  {Rosenbaum}}, \bibinfo {author} {\bibfnamefont {E.~D.}\ \bibnamefont
  {Isaacs}}, \bibinfo {author} {\bibfnamefont {O.~G.}\ \bibnamefont {Shpyrko}},
  \bibinfo {author} {\bibfnamefont {P.~G.}\ \bibnamefont {Evans}}, \bibinfo
  {author} {\bibfnamefont {G.}~\bibnamefont {Aeppli}}, \ and\ \bibinfo {author}
  {\bibfnamefont {Z.}~\bibnamefont {Cai}},\ }\href {\doibase
  10.1103/PhysRevLett.98.117206} {\bibfield  {journal} {\bibinfo  {journal}
  {Phys. Rev. Lett.}\ }\textbf {\bibinfo {volume} {98}},\ \bibinfo {pages}
  {117206} (\bibinfo {year} {2007})}\BibitemShut {NoStop}%
\bibitem [{\citenamefont {Sengupta}\ \emph {et~al.}(2010)\citenamefont
  {Sengupta}, \citenamefont {Solanki}, \citenamefont {Singh}, \citenamefont
  {Dhara},\ and\ \citenamefont {Deshmukh}}]{PhysRevB.82.155432}%
  \BibitemOpen
  \bibfield  {author} {\bibinfo {author} {\bibfnamefont {S.}~\bibnamefont
  {Sengupta}}, \bibinfo {author} {\bibfnamefont {H.~S.}\ \bibnamefont
  {Solanki}}, \bibinfo {author} {\bibfnamefont {V.}~\bibnamefont {Singh}},
  \bibinfo {author} {\bibfnamefont {S.}~\bibnamefont {Dhara}}, \ and\ \bibinfo
  {author} {\bibfnamefont {M.~M.}\ \bibnamefont {Deshmukh}},\ }\href {\doibase
  10.1103/PhysRevB.82.155432} {\bibfield  {journal} {\bibinfo  {journal} {Phys.
  Rev. B}\ }\textbf {\bibinfo {volume} {82}},\ \bibinfo {pages} {155432}
  (\bibinfo {year} {2010})}\BibitemShut {NoStop}%
\bibitem [{\citenamefont {Rossnagel}\ \emph {et~al.}(2001)\citenamefont
  {Rossnagel}, \citenamefont {Seifarth}, \citenamefont {Kipp}, \citenamefont
  {Skibowski}, \citenamefont {Vo\ss{}}, \citenamefont {Kr\"uger}, \citenamefont
  {Mazur},\ and\ \citenamefont {Pollmann}}]{PhysRevB.64.235119}%
  \BibitemOpen
  \bibfield  {author} {\bibinfo {author} {\bibfnamefont {K.}~\bibnamefont
  {Rossnagel}}, \bibinfo {author} {\bibfnamefont {O.}~\bibnamefont {Seifarth}},
  \bibinfo {author} {\bibfnamefont {L.}~\bibnamefont {Kipp}}, \bibinfo {author}
  {\bibfnamefont {M.}~\bibnamefont {Skibowski}}, \bibinfo {author}
  {\bibfnamefont {D.}~\bibnamefont {Vo\ss{}}}, \bibinfo {author} {\bibfnamefont
  {P.}~\bibnamefont {Kr\"uger}}, \bibinfo {author} {\bibfnamefont
  {A.}~\bibnamefont {Mazur}}, \ and\ \bibinfo {author} {\bibfnamefont
  {J.}~\bibnamefont {Pollmann}},\ }\href {\doibase 10.1103/PhysRevB.64.235119}
  {\bibfield  {journal} {\bibinfo  {journal} {Phys. Rev. B}\ }\textbf {\bibinfo
  {volume} {64}},\ \bibinfo {pages} {235119} (\bibinfo {year}
  {2001})}\BibitemShut {NoStop}%
\bibitem [{Note1()}]{Note1}%
  \BibitemOpen
  \bibinfo {note} {\label {myfootnote} The wavevector of the CDW state is known
  to be ${\protect \bf Q}_{\nu } \approx 1/3$ ${\protect \bf G}^{\nu }_0$,
  where ${\protect \bf G}^{\nu }_0$ are the three reciprocal lattice vectors,
  and $\nu =1, 2, 3$ in the 3Q phase. In the 1Q phase, only one of the
  ${\protect \bf Q}_{\nu }$ values remain active along the CDW propagation
  direction (we take ${\protect \bf Q}_1$)}\BibitemShut {NoStop}%
\bibitem [{\citenamefont {Ghosh}\ \emph {et~al.}(2004)\citenamefont {Ghosh},
  \citenamefont {Kar}, \citenamefont {Bid},\ and\ \citenamefont
  {Raychaudhuri}}]{ghosh2004set}%
  \BibitemOpen
  \bibfield  {author} {\bibinfo {author} {\bibfnamefont {A.}~\bibnamefont
  {Ghosh}}, \bibinfo {author} {\bibfnamefont {S.}~\bibnamefont {Kar}}, \bibinfo
  {author} {\bibfnamefont {A.}~\bibnamefont {Bid}}, \ and\ \bibinfo {author}
  {\bibfnamefont {A.}~\bibnamefont {Raychaudhuri}},\ }\href@noop {} {\bibfield
  {journal} {\bibinfo  {journal} {arXiv preprint cond-mat/0402130}\ } (\bibinfo
  {year} {2004})}\BibitemShut {NoStop}%
\bibitem [{\citenamefont {Soumyanarayanan}\ \emph {et~al.}(2013)\citenamefont
  {Soumyanarayanan}, \citenamefont {Yee}, \citenamefont {He}, \citenamefont
  {van Wezel}, \citenamefont {Rahn}, \citenamefont {Rossnagel}, \citenamefont
  {Hudson}, \citenamefont {Norman},\ and\ \citenamefont
  {Hoffman}}]{soumyanarayanan2013quantum}%
  \BibitemOpen
  \bibfield  {author} {\bibinfo {author} {\bibfnamefont {A.}~\bibnamefont
  {Soumyanarayanan}}, \bibinfo {author} {\bibfnamefont {M.~M.}\ \bibnamefont
  {Yee}}, \bibinfo {author} {\bibfnamefont {Y.}~\bibnamefont {He}}, \bibinfo
  {author} {\bibfnamefont {J.}~\bibnamefont {van Wezel}}, \bibinfo {author}
  {\bibfnamefont {D.~J.}\ \bibnamefont {Rahn}}, \bibinfo {author}
  {\bibfnamefont {K.}~\bibnamefont {Rossnagel}}, \bibinfo {author}
  {\bibfnamefont {E.}~\bibnamefont {Hudson}}, \bibinfo {author} {\bibfnamefont
  {M.~R.}\ \bibnamefont {Norman}}, \ and\ \bibinfo {author} {\bibfnamefont
  {J.~E.}\ \bibnamefont {Hoffman}},\ }\href@noop {} {\bibfield  {journal}
  {\bibinfo  {journal} {Proceedings of the National Academy of Sciences}\
  }\textbf {\bibinfo {volume} {110}},\ \bibinfo {pages} {1623} (\bibinfo {year}
  {2013})}\BibitemShut {NoStop}%
\bibitem [{\citenamefont {Levita}\ \emph {et~al.}(2014)\citenamefont {Levita},
  \citenamefont {Cavaleiro}, \citenamefont {Molinari}, \citenamefont {Polcar},\
  and\ \citenamefont {Righi}}]{levita2014sliding}%
  \BibitemOpen
  \bibfield  {author} {\bibinfo {author} {\bibfnamefont {G.}~\bibnamefont
  {Levita}}, \bibinfo {author} {\bibfnamefont {A.}~\bibnamefont {Cavaleiro}},
  \bibinfo {author} {\bibfnamefont {E.}~\bibnamefont {Molinari}}, \bibinfo
  {author} {\bibfnamefont {T.}~\bibnamefont {Polcar}}, \ and\ \bibinfo {author}
  {\bibfnamefont {M.}~\bibnamefont {Righi}},\ }\href@noop {} {\bibfield
  {journal} {\bibinfo  {journal} {The Journal of Physical Chemistry C}\
  }\textbf {\bibinfo {volume} {118}},\ \bibinfo {pages} {13809} (\bibinfo
  {year} {2014})}\BibitemShut {NoStop}%
\bibitem [{\citenamefont {Nagapriya}\ \emph {et~al.}(2008)\citenamefont
  {Nagapriya}, \citenamefont {Goldbart}, \citenamefont {Kaplan-Ashiri},
  \citenamefont {Seifert}, \citenamefont {Tenne},\ and\ \citenamefont
  {Joselevich}}]{nagapriya2008torsional}%
  \BibitemOpen
  \bibfield  {author} {\bibinfo {author} {\bibfnamefont {K.}~\bibnamefont
  {Nagapriya}}, \bibinfo {author} {\bibfnamefont {O.}~\bibnamefont {Goldbart}},
  \bibinfo {author} {\bibfnamefont {I.}~\bibnamefont {Kaplan-Ashiri}}, \bibinfo
  {author} {\bibfnamefont {G.}~\bibnamefont {Seifert}}, \bibinfo {author}
  {\bibfnamefont {R.}~\bibnamefont {Tenne}}, \ and\ \bibinfo {author}
  {\bibfnamefont {E.}~\bibnamefont {Joselevich}},\ }\href@noop {} {\bibfield
  {journal} {\bibinfo  {journal} {Physical review letters}\ }\textbf {\bibinfo
  {volume} {101}},\ \bibinfo {pages} {195501} (\bibinfo {year}
  {2008})}\BibitemShut {NoStop}%
\bibitem [{\citenamefont {Shmeliov}\ \emph {et~al.}(2014)\citenamefont
  {Shmeliov}, \citenamefont {Shannon}, \citenamefont {Wang}, \citenamefont
  {Kim}, \citenamefont {Okunishi}, \citenamefont {Nellist}, \citenamefont
  {Dolui}, \citenamefont {Sanvito},\ and\ \citenamefont {Nicolosi}}]{shmeliov}%
  \BibitemOpen
  \bibfield  {author} {\bibinfo {author} {\bibfnamefont {A.}~\bibnamefont
  {Shmeliov}}, \bibinfo {author} {\bibfnamefont {M.}~\bibnamefont {Shannon}},
  \bibinfo {author} {\bibfnamefont {P.}~\bibnamefont {Wang}}, \bibinfo {author}
  {\bibfnamefont {J.~S.}\ \bibnamefont {Kim}}, \bibinfo {author} {\bibfnamefont
  {E.}~\bibnamefont {Okunishi}}, \bibinfo {author} {\bibfnamefont {P.~D.}\
  \bibnamefont {Nellist}}, \bibinfo {author} {\bibfnamefont {K.}~\bibnamefont
  {Dolui}}, \bibinfo {author} {\bibfnamefont {S.}~\bibnamefont {Sanvito}}, \
  and\ \bibinfo {author} {\bibfnamefont {V.}~\bibnamefont {Nicolosi}},\
  }\href@noop {} {\bibfield  {journal} {\bibinfo  {journal} {ACS nano}\
  }\textbf {\bibinfo {volume} {8}},\ \bibinfo {pages} {3690} (\bibinfo {year}
  {2014})}\BibitemShut {NoStop}%
\bibitem [{\citenamefont {Sengupta}\ \emph {et~al.}(2010)\citenamefont
  {Sengupta}, \citenamefont {Solanki}, \citenamefont {Singh}, \citenamefont
  {Dhara},\ and\ \citenamefont {Deshmukh}}]{PhysRevB.82.155432}%
  \BibitemOpen
  \bibfield  {author} {\bibinfo {author} {\bibfnamefont {S.}~\bibnamefont
  {Sengupta}}, \bibinfo {author} {\bibfnamefont {H.~S.}\ \bibnamefont
  {Solanki}}, \bibinfo {author} {\bibfnamefont {V.}~\bibnamefont {Singh}},
  \bibinfo {author} {\bibfnamefont {S.}~\bibnamefont {Dhara}}, \ and\ \bibinfo
  {author} {\bibfnamefont {M.~M.}\ \bibnamefont {Deshmukh}},\ }\href {\doibase
  10.1103/PhysRevB.82.155432} {\bibfield  {journal} {\bibinfo  {journal} {Phys.
  Rev. B}\ }\textbf {\bibinfo {volume} {82}},\ \bibinfo {pages} {155432}
  (\bibinfo {year} {2010})}\BibitemShut {NoStop}%
\bibitem [{\citenamefont {Perdew}\ \emph {et~al.}(1996)\citenamefont {Perdew},
  \citenamefont {Burke},\ and\ \citenamefont {Ernzerhof}}]{PBE}%
  \BibitemOpen
  \bibfield  {author} {\bibinfo {author} {\bibfnamefont {J.~P.}\ \bibnamefont
  {Perdew}}, \bibinfo {author} {\bibfnamefont {K.}~\bibnamefont {Burke}}, \
  and\ \bibinfo {author} {\bibfnamefont {M.}~\bibnamefont {Ernzerhof}},\
  }\href@noop {} {\bibfield  {journal} {\bibinfo  {journal} {Physical review
  letters}\ }\textbf {\bibinfo {volume} {77}},\ \bibinfo {pages} {3865}
  (\bibinfo {year} {1996})}\BibitemShut {NoStop}%
\bibitem [{\citenamefont {Kresse}\ and\ \citenamefont
  {Furthm{\"u}ller}(1996)}]{vasp}%
  \BibitemOpen
  \bibfield  {author} {\bibinfo {author} {\bibfnamefont {G.}~\bibnamefont
  {Kresse}}\ and\ \bibinfo {author} {\bibfnamefont {J.}~\bibnamefont
  {Furthm{\"u}ller}},\ }\href@noop {} {\bibfield  {journal} {\bibinfo
  {journal} {Physical review B}\ }\textbf {\bibinfo {volume} {54}},\ \bibinfo
  {pages} {11169} (\bibinfo {year} {1996})}\BibitemShut {NoStop}%
\bibitem [{\citenamefont {Kresse}\ and\ \citenamefont {Joubert}(1999)}]{paw}%
  \BibitemOpen
  \bibfield  {author} {\bibinfo {author} {\bibfnamefont {G.}~\bibnamefont
  {Kresse}}\ and\ \bibinfo {author} {\bibfnamefont {D.}~\bibnamefont
  {Joubert}},\ }\href@noop {} {\bibfield  {journal} {\bibinfo  {journal}
  {Physical Review B}\ }\textbf {\bibinfo {volume} {59}},\ \bibinfo {pages}
  {1758} (\bibinfo {year} {1999})}\BibitemShut {NoStop}%
\bibitem [{\citenamefont {Gonze}(1995)}]{dfpt}%
  \BibitemOpen
  \bibfield  {author} {\bibinfo {author} {\bibfnamefont {X.}~\bibnamefont
  {Gonze}},\ }\href@noop {} {\bibfield  {journal} {\bibinfo  {journal}
  {Physical Review A}\ }\textbf {\bibinfo {volume} {52}},\ \bibinfo {pages}
  {1086} (\bibinfo {year} {1995})}\BibitemShut {NoStop}%
\bibitem [{\citenamefont {Togo}\ \emph {et~al.}(2008)\citenamefont {Togo},
  \citenamefont {Oba},\ and\ \citenamefont {Tanaka}}]{phonopy}%
  \BibitemOpen
  \bibfield  {author} {\bibinfo {author} {\bibfnamefont {A.}~\bibnamefont
  {Togo}}, \bibinfo {author} {\bibfnamefont {F.}~\bibnamefont {Oba}}, \ and\
  \bibinfo {author} {\bibfnamefont {I.}~\bibnamefont {Tanaka}},\ }\href@noop {}
  {\bibfield  {journal} {\bibinfo  {journal} {Physical Review B}\ }\textbf
  {\bibinfo {volume} {78}},\ \bibinfo {pages} {134106} (\bibinfo {year}
  {2008})}\BibitemShut {NoStop}%
\end{thebibliography}


Control: key (0)
Control: author (8) initials jnrlst
Control: editor formatted (1) identically to author
Control: production of article title (-1) disabled
Control: page (0) single
Control: year (1) truncated
Control: production of eprint (0) enabled

\newpage

\end{document}